\documentclass[aps,prl,superscriptaddress,showkeys,showpacs,twocolumn,longbibliography,nofootinbib]{revtex4-1}  
\usepackage[utf8]{inputenc}
\usepackage{amsmath}
\usepackage{amsfonts}
\usepackage{amsthm}
\usepackage{amssymb}
\usepackage{float}
\usepackage{braket}
\usepackage{graphicx}
\usepackage{url}
\usepackage[colorlinks=true, pdfstartview=FitV, linkcolor=red, citecolor=blue, urlcolor=blue]{hyperref}
\usepackage{slashed}
\usepackage[normalem]{ulem}

\begin{document}

\title{Onset of Bjorken Flow in Quantum Evolution of the Massive Schwinger Model}

\author{Haiyang Shao}
\email[]{shy21@mails.tsinghua.edu.cn}
\affiliation{Department of Physics, Tsinghua University, Beijing 100084, China.}

\author{Shile Chen}
\email[]{shile_chen@163.com}
\affiliation{Department of Physics, Tsinghua University, Beijing 100084, China.}

\author{Shuzhe Shi}
\email[]{shuzhe-shi@tsinghua.edu.cn}
\affiliation{Department of Physics, Tsinghua University, Beijing 100084, China.}
\affiliation{State Key Laboratory of Low-Dimensional Quantum Physics, Tsinghua University,
Beijing 100084, China.}

\begin{abstract}
The onset of hydrodynamics in the hot medium created in relativistic heavy-ion collisions is a crucial theoretical question. A first-principle simulation requires a real-time, non-perturbative calculation of the quantum system. In this Letter, we perform such simulations using the tensor network method, which enables large-scale quantum many-body simulations by retaining only the most essential quantum states for collective behaviors. We focus on the massive Schwinger model, a low-dimensional analog of quantum chromodynamics (QCD), as they share important properties such as confinement and chiral symmetry breaking.
Starting from an initial state that puts a localized excitation atop the vacuum and mimics the energy deposition from colliding nuclei, we observe hydrodynamic behavior consistent with Bjorken flow in all relevant degrees of freedom: energy density, fluid velocity, and bulk pressure. The time scale for hydrodynamic onset aligns with the thermalization time of the quantum distribution function.
\end{abstract}
\maketitle

\emph{Introduction.}--- 
Quark-gluon plasma (QGP), a novel phase of matter predicted by quantum chromodynamics (QCD), is central to high-energy nuclear physics. 
QGP has been identified as a nearly perfect fluid with extremely small dissipation, leading to the remarkable success of hydrodynamic simulations in describing and predicting high-statistics, precision measurements at the Relativistic Heavy-Ion Collider (RHIC) and the Large Hadron Collider (LHC)~\cite{Shuryak:2014zxa, Shen:2014vra, Schenke:2010rr, Karpenko:2013wva, vanderSchee:2013pia, Pang:2018zzo, Du:2019obx}. 
Assuming charge neutrality—valid for top-energy heavy-ion collisions—hydrodynamics is based on energy-momentum conservation ($\partial_\mu T^{\mu\nu}=0$) and thermodynamic properties. The stress tensor is decomposed as (e.g.,~\cite{1959flme.book.....L}),
\begin{align}
    T^{\mu\nu} = \varepsilon\, u^\mu u^\nu - \big(P+\Pi\big)(g^{\mu\nu} - u^\mu u^\nu)
    + \pi^{\mu\nu},
    \label{eq:Tmunu}
\end{align}
where $g^{\mu\nu}$ is the metric, $u^\mu$ the fluid velocity, while pressure $P$ and energy density $\varepsilon$ are related by the equation of state. Bulk pressure $\Pi$ and shear viscous tensor $\pi^{\mu\nu}$ are non-equilibrium dissipative corrections.
Hydrodynamics requires $\Pi$ and $\pi^{\mu\nu}$ to be small compared to $\varepsilon$ and $P$, implying proximity to local thermal equilibrium.

Given the highly non-thermal initial state of heavy-ion collisions, QGP fluidity presents a theoretical challenge. Previous approaches operate at classical or semi-classical levels: classical hydrodynamic analyses identify universal attractors emerging rapidly even from far-from-equilibrium states~\cite{Heller:2015dha, Heller:2013fn, Romatschke:2017vte, Romatschke:2017acs, Blaizot:2017ucy, Spalinski:2017mel, Ambrus:2021sjg, Kurkela:2019set, Almaalol:2020rnu, Blaizot:2019scw, Blaizot:2020gql, Blaizot:2021cdv, Chen:2024pez}; semi-classical treatments consider partonic degrees of freedom, studying thermalization via quark-gluon scatterings in Boltzmann equations~\cite{Baier:2000sb,Kurkela:2015qoa,Blaizot:2011xf, Blaizot:2013lga, Berges:2012us, Berges:2014bba, Blaizot:2014jna, Xu:2014ega, Martinez:2010sc, Martinez:2012tu, Ryblewski:2012rr, Strickland:2017kux, Strickland:2018ayk, Romatschke:2017ejr, Berges:2020fwq, Lu:2025yry}. 
Recent reviews on out-of-equilibrium hydrodynamics and QGP thermalization include Refs.~\cite{Romatschke:2017ejr, Alqahtani:2017mhy, Florkowski:2017olj, Berges:2020fwq, Shen:2020mgh, Soloviev:2021lhs, Jankowski:2023fdz, Strickland:2024moq}.

High-energy collisions produce strongly coupled, isolated quantum many-body systems evolving unitarily from pure initial states. How does a quantum system far from equilibrium spontaneously organize itself into a fluid described by hydrodynamics? This profound question lies at the heart of understanding QGP formation in heavy-ion collisions. While classical attractors and semi-classical Boltzmann approaches provide valuable insights, a fully quantum mechanical description of thermalization from first-principle QCD Hamiltonian remains elusive due to computational complexity.

As a QCD proxy, the Schwinger model~\cite{Schwinger:1962tp} shares essential features like confinement and chiral symmetry breaking~\cite{Lowenstein:1971fc, Jayewardena:1988td, Smilga:1992hx, Adam:1993fc, Adam:1997wt, Coleman:1976uz, Adam:1995us, Adam:1996np, Adam:1996qk}. Significant progress has been made in lattice studies using quantum~\cite{Klco:2018kyo, Farrell:2023fgd, Farrell:2024fit} and classical~\cite{Zache:2018cqq, Ikeda:2020agk, Kharzeev:2020kgc, deJong:2021wsd, Xie:2022jgj, Belyansky:2023rgh, Florio:2023dke, Florio:2024aix, Florio:2025hoc, Florio:2023mzk, Barata:2023jgd, Ikeda:2023zil, Ikeda:2023vfk, Lee:2023urk, Ghim:2024pxe} simulations, alongside other field theories (e.g.,~\cite{Li:2021kcs, Li:2022lyt, Li:2023kex, Czajka:2021yll, Carena:2022kpg, Yao:2023pht, Ebner:2023ixq, Ikeda:2024rzv, Hayata:2023puo, Hayata:2023pkw, Hidaka:2024zkd, Wu:2024adk, Carena:2024dzu, Qian:2024gph}). 
Particularly, in strong-coupling Schwinger model, thermal equilibrium emergence~\cite{Florio:2024aix, Florio:2025hoc} and collective motion~\cite{Janik:2025bbz} were studied in the presence of external sources, and thermalization of the Wigner function--the quantum analog of phase-space distributions--was observed in an isolated quantum system~\cite{Chen:2024pee}.
Stress tensor diffusion was also investigated in $\mathrm{SU}(2)$ lattice gauge theory~\cite{Turro:2025sec}.
See Refs.~\cite{Bauer:2022hpo, Bauer:2023qgm} for recent quantum simulation reviews in high-energy physics. 

In this work, we focus on the Schwinger model and address the onset of hydrodynamic behavior. Crucially, since hydrodynamics is fundamentally encoded in the evolution of the energy-momentum tensor, we directly probe hydrodynamic emergence by computing the full space-time evolution of $T^{\mu\nu}$ via quantum many-body simulations. We focus on an isolated quantum system starting from a local excitation on top of the vacuum state, avoiding complexity in the understanding due to the external sources. This approach also provides unambiguous access to:
\begin{enumerate}
    \item Local thermalization through $P(\varepsilon)$ convergence,
    \item Flow patterns via $u^\mu$ profiles,
    \item Dissipative dynamics from $\Pi$ evolution,
\end{enumerate}
allowing us to establish hydrodynamic onset without phenomenological assumptions. 
Studying hydrodynamics requires as many space grids as possible, which induces computational complexity because of the exponential scaling of Hilbert space. Thus, we use tensor networks~\cite{Rommer:1997zz, Buyens:2013yza, Buyens:2016ecr} in this work, which overcome the difficulty by dynamically retaining only entangled states relevant to real-time evolution.

\begin{figure}
    \centering
    \includegraphics[width=0.5\textwidth]{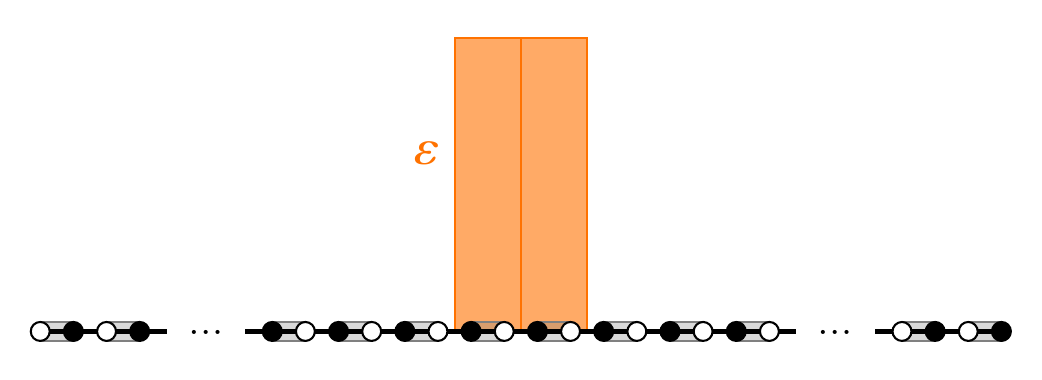}
    \caption{Staggered fermion (circles) and gauge link (lines) setup, with initial energy density (orange).
    \label{fig:illustration}}
\end{figure}
\vspace{5mm}
\emph{Model.}--- 
In the Schwinger model, interactions among fermion, antifermion, and an Abelian gauge field are characterized by the Action
\begin{align}
    \hat{S} = \int 
    \Big( \bar{\psi}\big(\gamma^\mu (i\partial_\mu - g\, A_\mu)- m \big)\psi - \frac{{F}^{\mu\nu} {F}_{\mu\nu}}{4}\Big)\mathrm{d}^2x\,,
    \label{eq:action}
\end{align}
in which $\psi$ and $\bar\psi$ are fermion field operators, $F^{\mu\nu}=\partial^\mu A^\nu - \partial^\nu A^\mu$ gives the gauge field strength, $m$ is the fermion (bare) mass and $g$ the coupling constant.  
To mimic the feature of quark-gluon plasma, we will consider these fermions as one-flavor quarks, and the coupling $g$ as the strong coupling constant.
Notably, $m$ and $g$, both of the energy unit, are the only two parameters in the theory. We set $g$ as the energy unit and explore the effect of coupling and/or mass by varying the ratio ($m/g$).
We employ the natural unit $\hbar=c=k_B=1$ throughout this paper.

The stress tensor of the Schwinger model can be obtained by taking the functional derivative over the Action~\eqref{eq:action} with respect to the metric, $\hat{T}^{\mu\nu} \equiv\frac{\delta \hat S}{\delta g_{\mu\nu}}$, which gives that
\begin{align}
\begin{split}
\hat{T}^{tt} 
=\;& 
    \frac{\mathcal{E}^2}{2} 
    + \frac{i}{2} (\bar{\psi}' \gamma^1 \psi
    - \bar{\psi} \gamma^1 \psi')
    + \bar{\psi}(g\gamma^1 A_1 + m)\psi\,,\\
\hat{T}^{zz} 
=\;&
    -\frac{\mathcal{E}^2}{2} 
    + \frac{i}{2} (\bar{\psi}' \gamma^1 \psi
    - \bar{\psi} \gamma^1 \psi')
    + g \bar{\psi} \gamma^1 A_1\psi
\,,\\
\hat{T}^{zt} 
=\;&
\hat{T}^{tz}
=
     \frac{i}{2} (\psi'^\dagger \psi
     - \psi^\dagger \psi')
     + g\,A_1\, \psi^\dagger \psi,
\end{split}\label{eq:stress_tensor}
\end{align}
where $\bar{\psi}' \equiv \partial_z\bar{\psi}$ and likewise for ${\psi}'$, $A_1$ is the gauge field potential, which gives rise to the electric field $\mathcal{E}=\partial_0 A_1$. In particular, $\hat{T}^{tt}$ gives the energy density of the system, which gives the Hamiltonian $\hat{H} \equiv \int \hat{T}^{tt} \mathrm{d}z$. 
We have taken Gauss' law, $\partial_z \mathcal{E} - g\, \psi^\dagger \psi = 0$, to ensure gauge invariance. The Dirac equation, $\partial_t \psi \equiv i[\hat H, \psi]$, is also implemented so that one can take the Schr\"odinger picture and construct time-independent operators. With an initial state $\ket{\Psi(0)}$, the time-dependent quantum state is given by 
\begin{align}
    \ket{\Psi(t)} = e^{-i\hat{H}t}\ket{\Psi(0)}\,.
    \label{eq:schroedinger}
\end{align}
Noting that $i[\hat{H}, \hat{T}^{t\nu}] = -\partial_z \hat{T}^{z\nu}$ for $\nu \in \{t,z\}$, energy-momentum conservation holds at the operator level.

\begin{figure}
    \centering
    \includegraphics[width=0.5\textwidth]{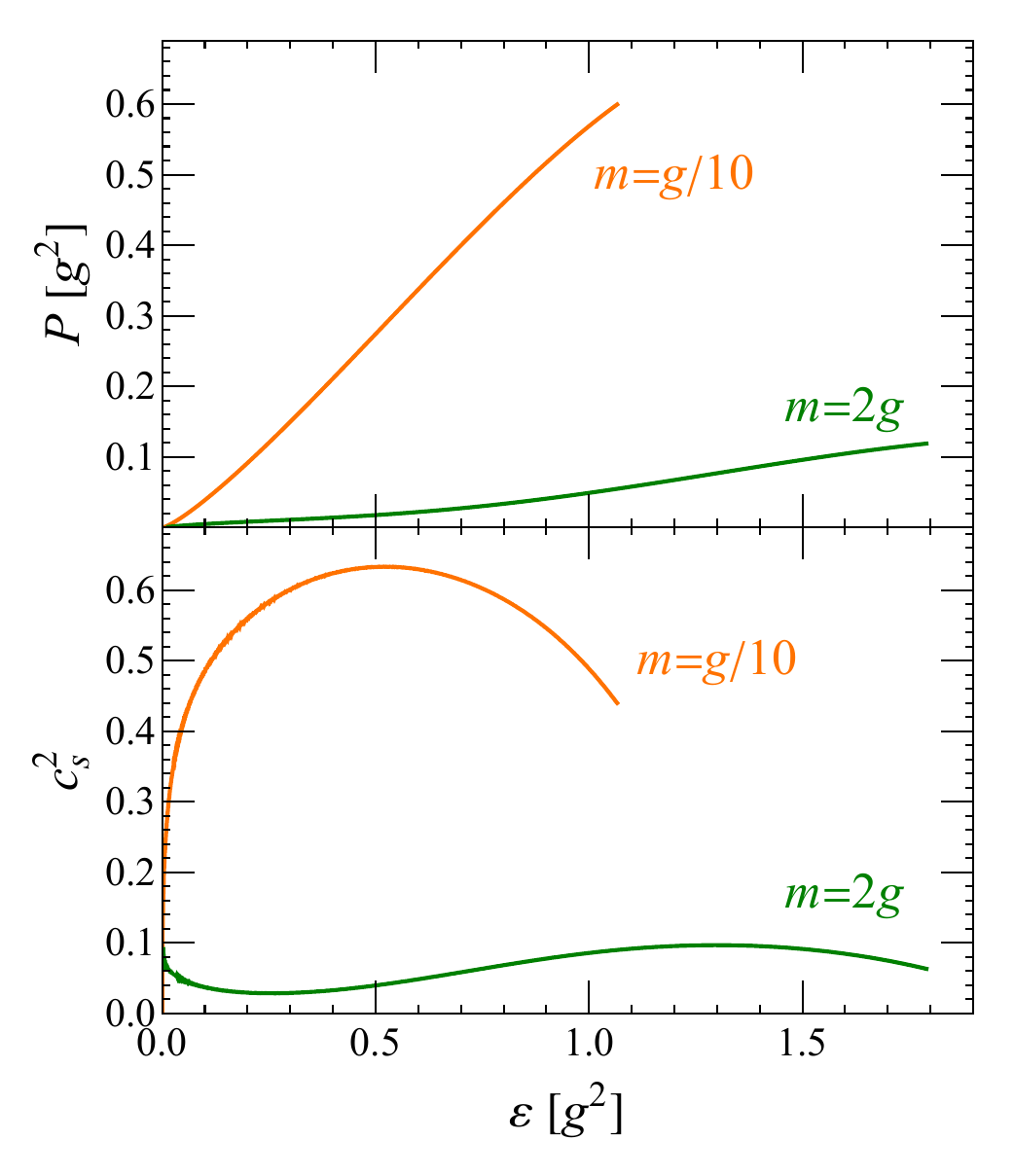}
    \caption{Pressure (top) and speed of sound (bottom) versus energy density for $m=g/10$ (orange) and $m=2g$ (green).
    \label{fig:eos}}
\end{figure}

Numerical simulations of quantum field theories must be performed in finite volume and discretized (coordinate) space.
We consider $z \in [0, L]$ with open boundaries $\psi(0) = \psi(L) = 0$, $\mathcal{E}(0) = \mathcal{E}(L) = 0$. We discretize the coordinate into $N=100$ lattice sites and take the Kogut--Susskind~\cite{Kogut:1974ag, Susskind:1976jm} staggered fermion representation which puts the fermion (anti-fermion) on the even (odd) sites (see Fig.~\ref{fig:illustration} for illustration). The lattice spacing is set to $a\equiv L/N = 1/2\,g^{-1}$ in which thermalization of the Schwinger model has been observed~\cite{Brenes:2017wzd, Florio:2024aix}. We represent quantum states by the Matrix Product States representation of tensor network~\cite{Rommer:1997zz, Buyens:2013yza, Buyens:2016ecr} with bond dimension $\mathcal{D}=500$. To mimic the hot medium created in heavy-ion collisions, the initial quantum state is prepared as the vacuum (i.e., ground state of the Hamiltonian) perturbed by a local excitation of local energy density, [$\varphi(z) = \pi\,\Theta(-z+a/2)\Theta(z+a/2)$ with $\Theta$ being the step function]
\begin{align}
    \ket{\Psi(0)} = e^{i \int \varphi(z) \bar\psi(z) \psi(z) dz} \ket{\mathrm{vac}},
\end{align}
See Fig.~\ref{fig:illustration} for illustration. 
With the initial condition, the time dependent wavefunction~\eqref{eq:schroedinger} is solved by Time-Evolving Block Decimation method~\cite{Vidal:2003lvx} with time step size $\Delta t = 0.01\, g^{-1}$.

\begin{figure}[!hbt]
    \centering
    \includegraphics[width=0.5\textwidth]{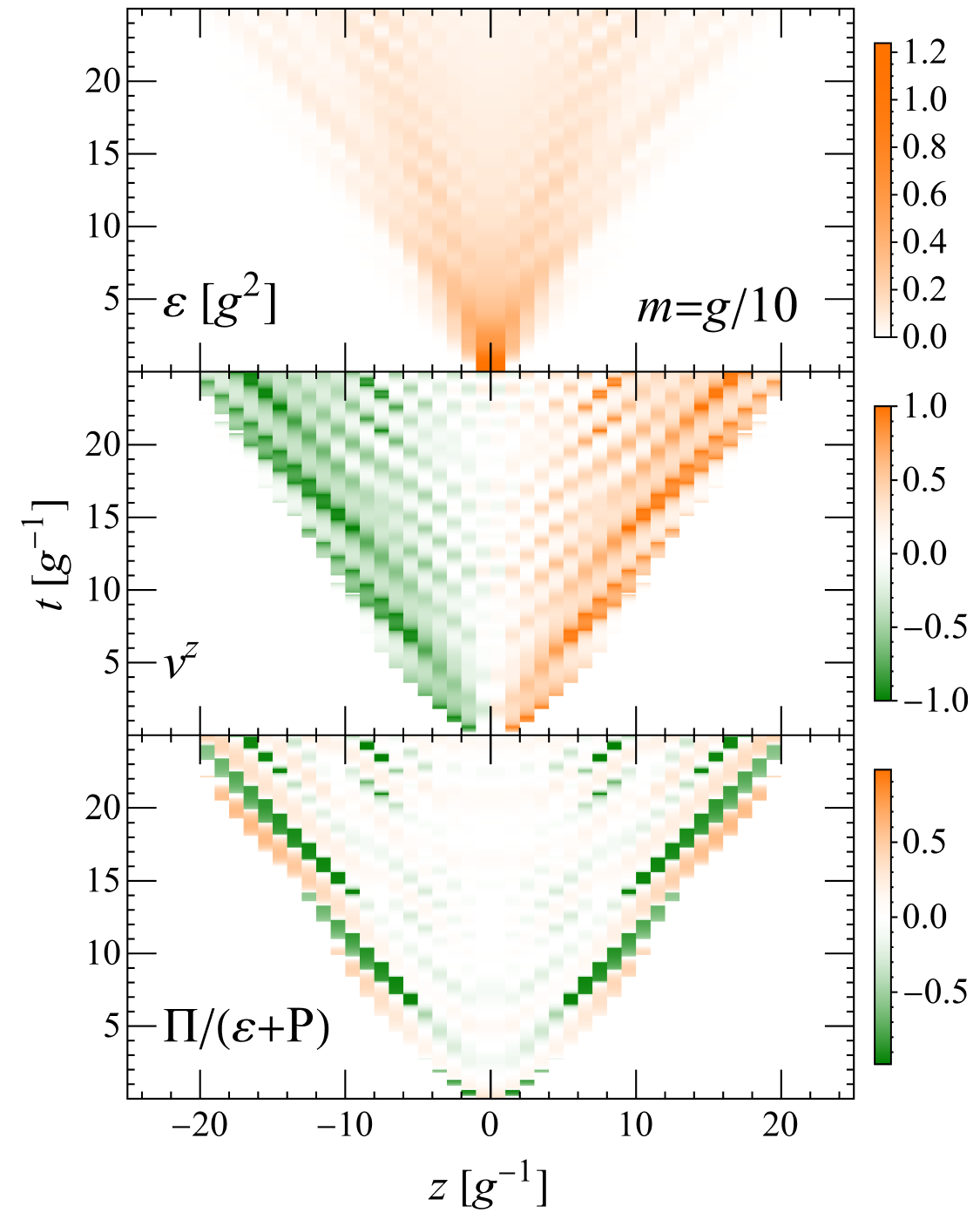}
    \caption{Energy density (top), fluid velocity (middle), and $\Pi/(\varepsilon+P)$ (bottom) for $m=g/10$.
    \label{fig:lightcone}}
\end{figure}
Elements in the stress tensor are given by the expectation values of the corresponding operators modulo by the vacuum expectations,\footnote{
In non-equilibrium systems, definition of fluid velocity is not unique, and we have taken the Landau--Lifshitz frame~\cite{1959flme.book.....L} which fixes it according to the energy flow, $T^{\mu\nu} u_\nu = \varepsilon\, u^\mu$.}
\begin{align}
    T^{\mu\nu}(t,z) \equiv \bra{\Psi(t)} \hat{T}^{\mu\nu}(z) \ket{\Psi(t)} - \bra{\mathrm{vac}} \hat{T}^{\mu\nu}(z) \ket{\mathrm{vac}}\,.
    \label{eq:stress_tensor}
\end{align}
In $1+1$ D, the stress tensor has only four elements. There is no shear viscous tensor ($\pi^{\mu\nu}$), and the bulk pressure measures the difference between the effective pressure and the one determined by the equation of state (EoS), $P = P(\varepsilon)$. We compute the EoS via thermodynamic relations, $P(T) = \frac{\mathrm{tr}(\hat{T}^{zz} e^{-\hat{H}/T})}{\mathrm{tr}(e^{-\hat{H}/T})}$ and $\varepsilon(T) = \frac{\mathrm{tr}(\hat{T}^{tt} e^{-\hat{H}/T})}{\mathrm{tr}(e^{-\hat{H}/T})}$, as shown in Fig.~\ref{fig:eos}. 
Details of the numerical method---including discretization, operator representation, initial state preparation, real time evolution, and finite temperature calculations---can be referred to our companion paper~\cite{Shao_2025_long}.

\vspace{5mm}
\emph{Hydrodynamic Evolution.}--- 
The space-time profiles of energy density, fluid velocity, and the bulk-pressure-to-enthalpy ratio are shown in Fig.~\ref{fig:lightcone}, which clearly exhibit the medium expansion within light-cone, with a velocity that increases with the distance from the origin. Particularly, the bulk pressure becomes much smaller compared to the enthalpy as time increases. For sufficiently long time, the effective pressure in the dynamical evolution ($P+\Pi$) is in good agreement with the thermal one [$P(\varepsilon)$], indicating that the system has reached local thermal equilibrium as the viscous corrections are negligible.

\begin{figure}[!hbt]
    \centering
    \includegraphics[width=0.45\textwidth]{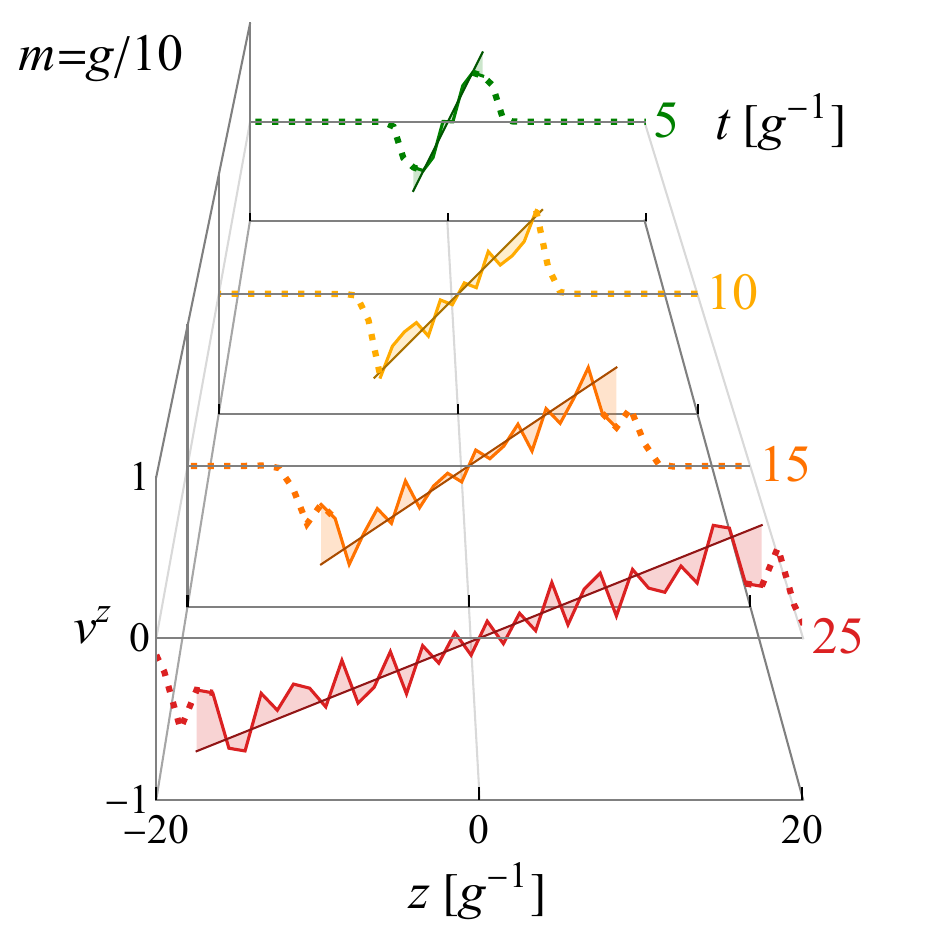}
    \caption{Fluid velocity for $m=g/10$. Curves: simulations; straight lines: $v^z = z/t$. Solid (dashed) curves are inside (outside) the lightcone.
    \label{fig:velocity}}
\end{figure}

All these characteristics resemble the Bjorken solution of ideal hydrodynamic equations with finite overlap time~\cite{Hwa:1974gn, Bjorken:1982qr, Shi:2022iyb}. The Bjorken flow assumes that the hot medium expands in a manner that is invariant under a Lorentz boost along the expansion direction. Its flow velocity reads $v^z = z/t$, and the energy density depends only on the proper time ($\tau\equiv\sqrt{t^2-z^2}$) but not the rapidity ($\eta \equiv \frac{1}{2}\ln\frac{t+z}{t-z}$), i.e., $\varepsilon \propto \tau^{-1-c_s^2}$.
The non-equilibrium correction can also be calculated in a boost-invariant system. In the Navier--Stokes theory which keeps the viscous correction up to the first order in gradient expansion, second law of thermodynamics requires that $\Pi = \zeta \,s\, \theta$~\cite{1959flme.book.....L}, with $s$ being the entropy density, $\theta\equiv \partial_\mu u^\mu$ the expansion rate, and the positive constant $\zeta$ is referred to as bulk viscosity. Taking the ideal solution for $s$ and noting that $\theta=1/\tau$, one may obtain the long-time behavior of the bulk pressure $\Pi \propto \tau^{-2}$.

\begin{figure}
    \centering
    \includegraphics[width=0.5\textwidth]{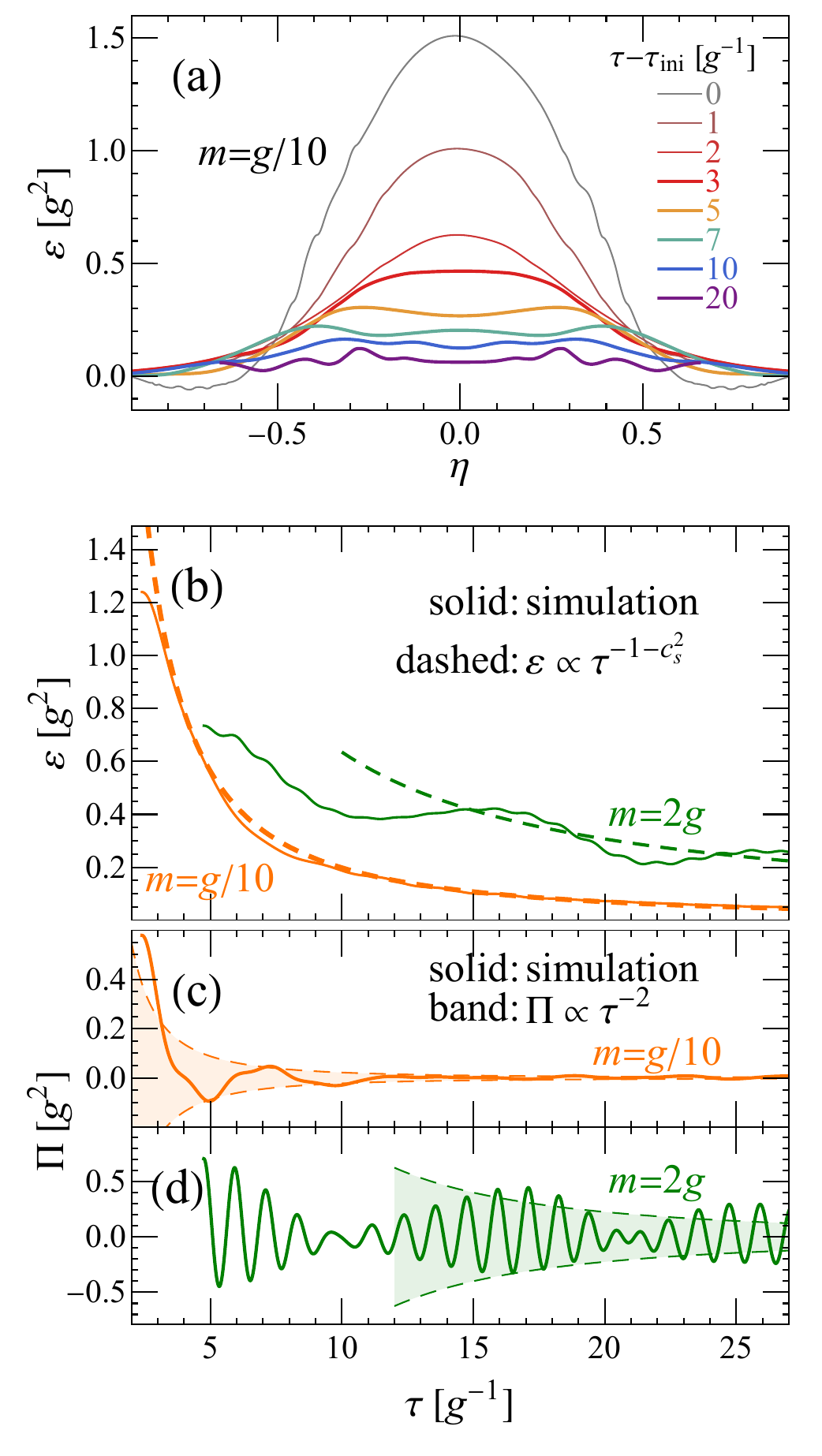}
    \caption{(a) Energy density rapidity distributions. (b-d) Proper time dependence of energy density (b) and bulk pressure (c,d). Solid curves: simulations for $m=g/10$ (orange), $m=2g$ (green); dashed curves: Bjorken solutions.
    \label{fig:hydro}}
\end{figure}

Fig.~\ref{fig:velocity} compares simulated flow velocity to Bjorken flow, showing agreement inside the lightcone despite $\mathcal{O}(0.1)$ quantum fluctuations. The generalized Bjorken flow also predicts a plateau structure of energy density at mid rapidity~\cite{Shi:2022iyb}, which is also observed in Fig.~\ref{fig:hydro}~(a), for the proper time $\tau - \tau_\mathrm{ini} \gtrsim 3 g^{-1}$; as time increases, the width of the rapidity plateau expands while the energy density decreases; at late time ($\tau - \tau_\mathrm{ini} \sim 20 g^{-1}$), the energy density becomes dilute, and the influence of quantum fluctuations becomes more pronounced especially in the large rapidity region. Noting that the current analysis is performed for a single, pure state, we anticipate that the effect of quantum fluctuations would be suppressed by taking the average over different initial states, which corresponds to taking the average over different events in heavy-ion collisions.

The proper time dependence of the energy density and bulk pressure at mid rapidity is also quantitatively consistent with the Bjorken flow in the strong coupling case. As shown in Fig.~\ref{fig:hydro}~(b), we observe that the result of the quantum simulation of $\varepsilon$ (orange solid curve) well resembles the Bjorken solution, $\varepsilon\propto \tau^{-1-c_s^2}$, (orange dashed curve). In the latter, the speed of sound $c_s^2 = 0.5$ is taken from the EoS (Fig.~\ref{fig:eos}) at the corresponding energy density.
Local thermal equilibrium---judged by the difference between simulation result and the Bjorken solution---is reached at a timescale $\tau \sim 10 g^{-1}$. This timescale is comparable with the thermalization time of the momentum distribution function in the Schwinger model~\cite{Chen:2024pee}.

Additionally, while the bulk pressure starts from a large value, it drops rapidly, oscillates around zero, and eventually vanishes [see Fig.~\ref{fig:hydro}~(c)]. After $\tau - \tau_\mathrm{ini} \gtrsim 1\,g^{-1}$, the amplitude of the oscillation is well bounded within the Navier--Stokes solution ($\Pi \propto \tau^{-2}$), represented by the orange band. Comparison of the time scales reveals hydrodynamization faster than global thermalization.
In contrast, in the case of large fermion mass/weak coupling, neither does the energy density resemble the Bjorken flow (with $c_s^2=0.05$) nor does oscillation of bulk pressure diminish quickly, suggesting the absence of the Bjorken-like hydrodynamic behavior.

\vspace{5mm}
\emph{Summary and Discussion.}--- In this Letter, we studied the onset of hydrodynamic behaviors in isolated quantum systems by performing quantum many-body simulation for the massive Schwinger model. With the initial state---a localized excitation atop the vacuum---resembling the energy deposition from colliding nuclei where longitudinal fields dominate over transverse expansion in early stages, we observe the onset of hydrodynamic behavior that resembles the generalized Bjorken flow with finite collision time, in the case of a strong coupling constant, or equivalently a small fermion mass. Good quantitative consistency is observed in the fluid velocity profile, rapidity plateau structure in the energy distribution, and the time dependence of the energy density and bulk pressure. 

Our simulation also suggests that the Bjorken-like hydrodynamic behavior is absent when the coupling is weak, or equivalently, the fermions are heavy. Such a contrast sheds light on the key feature in the underlying microscopic theory that is crucial to the rapid development of the hydrodynamic behaviors. Intuitively, a strong coupling constant enhances particle scattering, and a small fermion mass further promotes the pair production and annihilation processes in microscopic interactions---both of them accelerate the thermalization process. To address the question that which of them plays a more important role, one needs to perform the same analysis on a theory in which the role of mass and coupling can be separated. Results will be reported in our future publication.

\section*{Acknowledgments}
We thank Sangyong Jeon, Weiyao Ke, Tongyu Lin, Dmitri Kharzeev, Li Yan, Yi Yin, Yuanchen Zhao, and Pengfei Zhuang for discussions. This work is supported by NSFC under grant No. 12575143 and Tsinghua University under grants No. 04200500123, No. 531205006, and No. 533305009.
Computations are performed at Center of High performance computing, Tsinghua University.
\bibliography{ref}

\begin{thebibliography}{101}%
\makeatletter
\providecommand \@ifxundefined [1]{%
 \@ifx{#1\undefined}
}%
\providecommand \@ifnum [1]{%
 \ifnum #1\expandafter \@firstoftwo
 \else \expandafter \@secondoftwo
 \fi
}%
\providecommand \@ifx [1]{%
 \ifx #1\expandafter \@firstoftwo
 \else \expandafter \@secondoftwo
 \fi
}%
\providecommand \natexlab [1]{#1}%
\providecommand \enquote  [1]{``#1''}%
\providecommand \bibnamefont  [1]{#1}%
\providecommand \bibfnamefont [1]{#1}%
\providecommand \citenamefont [1]{#1}%
\providecommand \href@noop [0]{\@secondoftwo}%
\providecommand \href [0]{\begingroup \@sanitize@url \@href}%
\providecommand \@href[1]{\@@startlink{#1}\@@href}%
\providecommand \@@href[1]{\endgroup#1\@@endlink}%
\providecommand \@sanitize@url [0]{\catcode `\\12\catcode `\$12\catcode
  `\&12\catcode `\#12\catcode `\^12\catcode `\_12\catcode `\%12\relax}%
\providecommand \@@startlink[1]{}%
\providecommand \@@endlink[0]{}%
\providecommand \url  [0]{\begingroup\@sanitize@url \@url }%
\providecommand \@url [1]{\endgroup\@href {#1}{\urlprefix }}%
\providecommand \urlprefix  [0]{URL }%
\providecommand \Eprint [0]{\href }%
\providecommand \doibase [0]{http://dx.doi.org/}%
\providecommand \selectlanguage [0]{\@gobble}%
\providecommand \bibinfo  [0]{\@secondoftwo}%
\providecommand \bibfield  [0]{\@secondoftwo}%
\providecommand \translation [1]{[#1]}%
\providecommand \BibitemOpen [0]{}%
\providecommand \bibitemStop [0]{}%
\providecommand \bibitemNoStop [0]{.\EOS\space}%
\providecommand \EOS [0]{\spacefactor3000\relax}%
\providecommand \BibitemShut  [1]{\csname bibitem#1\endcsname}%
\let\auto@bib@innerbib\@empty
%</preamble>
\bibitem [{\citenamefont {Shuryak}(2017)}]{Shuryak:2014zxa}%
  \BibitemOpen
  \bibfield  {author} {\bibinfo {author} {\bibfnamefont {Edward}\ \bibnamefont
  {Shuryak}},\ }\bibfield  {title} {\enquote {\bibinfo {title} {{Strongly
  coupled quark-gluon plasma in heavy ion collisions}},}\ }\href {\doibase
  10.1103/RevModPhys.89.035001} {\bibfield  {journal} {\bibinfo  {journal}
  {Rev. Mod. Phys.}\ }\textbf {\bibinfo {volume} {89}},\ \bibinfo {pages}
  {035001} (\bibinfo {year} {2017})},\ \Eprint {http://arxiv.org/abs/1412.8393}
  {arXiv:1412.8393 [hep-ph]} \BibitemShut {NoStop}%
\bibitem [{\citenamefont {Shen}\ \emph {et~al.}(2016)\citenamefont {Shen},
  \citenamefont {Qiu}, \citenamefont {Song}, \citenamefont {Bernhard},
  \citenamefont {Bass},\ and\ \citenamefont {Heinz}}]{Shen:2014vra}%
  \BibitemOpen
  \bibfield  {author} {\bibinfo {author} {\bibfnamefont {Chun}\ \bibnamefont
  {Shen}}, \bibinfo {author} {\bibfnamefont {Zhi}\ \bibnamefont {Qiu}},
  \bibinfo {author} {\bibfnamefont {Huichao}\ \bibnamefont {Song}}, \bibinfo
  {author} {\bibfnamefont {Jonah}\ \bibnamefont {Bernhard}}, \bibinfo {author}
  {\bibfnamefont {Steffen}\ \bibnamefont {Bass}}, \ and\ \bibinfo {author}
  {\bibfnamefont {Ulrich}\ \bibnamefont {Heinz}},\ }\bibfield  {title}
  {\enquote {\bibinfo {title} {{The iEBE-VISHNU code package for relativistic
  heavy-ion collisions}},}\ }\href {\doibase 10.1016/j.cpc.2015.08.039}
  {\bibfield  {journal} {\bibinfo  {journal} {Comput. Phys. Commun.}\ }\textbf
  {\bibinfo {volume} {199}},\ \bibinfo {pages} {61--85} (\bibinfo {year}
  {2016})},\ \Eprint {http://arxiv.org/abs/1409.8164} {arXiv:1409.8164
  [nucl-th]} \BibitemShut {NoStop}%
\bibitem [{\citenamefont {Schenke}\ \emph {et~al.}(2011)\citenamefont
  {Schenke}, \citenamefont {Jeon},\ and\ \citenamefont
  {Gale}}]{Schenke:2010rr}%
  \BibitemOpen
  \bibfield  {author} {\bibinfo {author} {\bibfnamefont {Bjorn}\ \bibnamefont
  {Schenke}}, \bibinfo {author} {\bibfnamefont {Sangyong}\ \bibnamefont
  {Jeon}}, \ and\ \bibinfo {author} {\bibfnamefont {Charles}\ \bibnamefont
  {Gale}},\ }\bibfield  {title} {\enquote {\bibinfo {title} {{Elliptic and
  triangular flow in event-by-event (3+1)D viscous hydrodynamics}},}\ }\href
  {\doibase 10.1103/PhysRevLett.106.042301} {\bibfield  {journal} {\bibinfo
  {journal} {Phys. Rev. Lett.}\ }\textbf {\bibinfo {volume} {106}},\ \bibinfo
  {pages} {042301} (\bibinfo {year} {2011})},\ \Eprint
  {http://arxiv.org/abs/1009.3244} {arXiv:1009.3244 [hep-ph]} \BibitemShut
  {NoStop}%
\bibitem [{\citenamefont {Karpenko}\ \emph {et~al.}(2014)\citenamefont
  {Karpenko}, \citenamefont {Huovinen},\ and\ \citenamefont
  {Bleicher}}]{Karpenko:2013wva}%
  \BibitemOpen
  \bibfield  {author} {\bibinfo {author} {\bibfnamefont {Iu.}\ \bibnamefont
  {Karpenko}}, \bibinfo {author} {\bibfnamefont {P.}~\bibnamefont {Huovinen}},
  \ and\ \bibinfo {author} {\bibfnamefont {M.}~\bibnamefont {Bleicher}},\
  }\bibfield  {title} {\enquote {\bibinfo {title} {{A 3+1 dimensional viscous
  hydrodynamic code for relativistic heavy ion collisions}},}\ }\href {\doibase
  10.1016/j.cpc.2014.07.010} {\bibfield  {journal} {\bibinfo  {journal}
  {Comput. Phys. Commun.}\ }\textbf {\bibinfo {volume} {185}},\ \bibinfo
  {pages} {3016--3027} (\bibinfo {year} {2014})},\ \Eprint
  {http://arxiv.org/abs/1312.4160} {arXiv:1312.4160 [nucl-th]} \BibitemShut
  {NoStop}%
\bibitem [{\citenamefont {van~der Schee}\ \emph {et~al.}(2013)\citenamefont
  {van~der Schee}, \citenamefont {Romatschke},\ and\ \citenamefont
  {Pratt}}]{vanderSchee:2013pia}%
  \BibitemOpen
  \bibfield  {author} {\bibinfo {author} {\bibfnamefont {Wilke}\ \bibnamefont
  {van~der Schee}}, \bibinfo {author} {\bibfnamefont {Paul}\ \bibnamefont
  {Romatschke}}, \ and\ \bibinfo {author} {\bibfnamefont {Scott}\ \bibnamefont
  {Pratt}},\ }\bibfield  {title} {\enquote {\bibinfo {title} {{Fully Dynamical
  Simulation of Central Nuclear Collisions}},}\ }\href {\doibase
  10.1103/PhysRevLett.111.222302} {\bibfield  {journal} {\bibinfo  {journal}
  {Phys. Rev. Lett.}\ }\textbf {\bibinfo {volume} {111}},\ \bibinfo {pages}
  {222302} (\bibinfo {year} {2013})},\ \Eprint {http://arxiv.org/abs/1307.2539}
  {arXiv:1307.2539 [nucl-th]} \BibitemShut {NoStop}%
\bibitem [{\citenamefont {Pang}\ \emph {et~al.}(2018)\citenamefont {Pang},
  \citenamefont {Petersen},\ and\ \citenamefont {Wang}}]{Pang:2018zzo}%
  \BibitemOpen
  \bibfield  {author} {\bibinfo {author} {\bibfnamefont {Long-Gang}\
  \bibnamefont {Pang}}, \bibinfo {author} {\bibfnamefont {Hannah}\ \bibnamefont
  {Petersen}}, \ and\ \bibinfo {author} {\bibfnamefont {Xin-Nian}\ \bibnamefont
  {Wang}},\ }\bibfield  {title} {\enquote {\bibinfo {title} {{Pseudorapidity
  distribution and decorrelation of anisotropic flow within the
  open-computing-language implementation CLVisc hydrodynamics}},}\ }\href
  {\doibase 10.1103/PhysRevC.97.064918} {\bibfield  {journal} {\bibinfo
  {journal} {Phys. Rev. C}\ }\textbf {\bibinfo {volume} {97}},\ \bibinfo
  {pages} {064918} (\bibinfo {year} {2018})},\ \Eprint
  {http://arxiv.org/abs/1802.04449} {arXiv:1802.04449 [nucl-th]} \BibitemShut
  {NoStop}%
\bibitem [{\citenamefont {Du}\ and\ \citenamefont {Heinz}(2020)}]{Du:2019obx}%
  \BibitemOpen
  \bibfield  {author} {\bibinfo {author} {\bibfnamefont {Lipei}\ \bibnamefont
  {Du}}\ and\ \bibinfo {author} {\bibfnamefont {Ulrich}\ \bibnamefont
  {Heinz}},\ }\bibfield  {title} {\enquote {\bibinfo {title}
  {{(3+1)-dimensional dissipative relativistic fluid dynamics at non-zero net
  baryon density}},}\ }\href {\doibase 10.1016/j.cpc.2019.107090} {\bibfield
  {journal} {\bibinfo  {journal} {Comput. Phys. Commun.}\ }\textbf {\bibinfo
  {volume} {251}},\ \bibinfo {pages} {107090} (\bibinfo {year} {2020})},\
  \Eprint {http://arxiv.org/abs/1906.11181} {arXiv:1906.11181 [nucl-th]}
  \BibitemShut {NoStop}%
\bibitem [{\citenamefont {{Landau}}\ and\ \citenamefont
  {{Lifshitz}}(1959)}]{1959flme.book.....L}%
  \BibitemOpen
  \bibfield  {author} {\bibinfo {author} {\bibfnamefont {Lev~Davidovich}\
  \bibnamefont {{Landau}}}\ and\ \bibinfo {author} {\bibfnamefont {E.~M.}\
  \bibnamefont {{Lifshitz}}},\ }\href@noop {} {\emph {\bibinfo {title} {{Fluid
  mechanics}}}}\ (\bibinfo {year} {1959})\BibitemShut {NoStop}%
\bibitem [{\citenamefont {Heller}\ and\ \citenamefont
  {Spalinski}(2015)}]{Heller:2015dha}%
  \BibitemOpen
  \bibfield  {author} {\bibinfo {author} {\bibfnamefont {Michal~P.}\
  \bibnamefont {Heller}}\ and\ \bibinfo {author} {\bibfnamefont {Michal}\
  \bibnamefont {Spalinski}},\ }\bibfield  {title} {\enquote {\bibinfo {title}
  {{Hydrodynamics Beyond the Gradient Expansion: Resurgence and
  Resummation}},}\ }\href {\doibase 10.1103/PhysRevLett.115.072501} {\bibfield
  {journal} {\bibinfo  {journal} {Phys. Rev. Lett.}\ }\textbf {\bibinfo
  {volume} {115}},\ \bibinfo {pages} {072501} (\bibinfo {year} {2015})},\
  \Eprint {http://arxiv.org/abs/1503.07514} {arXiv:1503.07514 [hep-th]}
  \BibitemShut {NoStop}%
\bibitem [{\citenamefont {Heller}\ \emph {et~al.}(2013)\citenamefont {Heller},
  \citenamefont {Janik},\ and\ \citenamefont {Witaszczyk}}]{Heller:2013fn}%
  \BibitemOpen
  \bibfield  {author} {\bibinfo {author} {\bibfnamefont {Michal~P.}\
  \bibnamefont {Heller}}, \bibinfo {author} {\bibfnamefont {Romuald~A.}\
  \bibnamefont {Janik}}, \ and\ \bibinfo {author} {\bibfnamefont {Przemyslaw}\
  \bibnamefont {Witaszczyk}},\ }\bibfield  {title} {\enquote {\bibinfo {title}
  {{Hydrodynamic Gradient Expansion in Gauge Theory Plasmas}},}\ }\href
  {\doibase 10.1103/PhysRevLett.110.211602} {\bibfield  {journal} {\bibinfo
  {journal} {Phys. Rev. Lett.}\ }\textbf {\bibinfo {volume} {110}},\ \bibinfo
  {pages} {211602} (\bibinfo {year} {2013})},\ \Eprint
  {http://arxiv.org/abs/1302.0697} {arXiv:1302.0697 [hep-th]} \BibitemShut
  {NoStop}%
\bibitem [{\citenamefont {Romatschke}(2018)}]{Romatschke:2017vte}%
  \BibitemOpen
  \bibfield  {author} {\bibinfo {author} {\bibfnamefont {Paul}\ \bibnamefont
  {Romatschke}},\ }\bibfield  {title} {\enquote {\bibinfo {title}
  {{Relativistic Fluid Dynamics Far From Local Equilibrium}},}\ }\href
  {\doibase 10.1103/PhysRevLett.120.012301} {\bibfield  {journal} {\bibinfo
  {journal} {Phys. Rev. Lett.}\ }\textbf {\bibinfo {volume} {120}},\ \bibinfo
  {pages} {012301} (\bibinfo {year} {2018})},\ \Eprint
  {http://arxiv.org/abs/1704.08699} {arXiv:1704.08699 [hep-th]} \BibitemShut
  {NoStop}%
\bibitem [{\citenamefont {Romatschke}(2017)}]{Romatschke:2017acs}%
  \BibitemOpen
  \bibfield  {author} {\bibinfo {author} {\bibfnamefont {Paul}\ \bibnamefont
  {Romatschke}},\ }\bibfield  {title} {\enquote {\bibinfo {title}
  {{Relativistic Hydrodynamic Attractors with Broken Symmetries: Non-Conformal
  and Non-Homogeneous}},}\ }\href {\doibase 10.1007/JHEP12(2017)079} {\bibfield
   {journal} {\bibinfo  {journal} {JHEP}\ }\textbf {\bibinfo {volume} {12}},\
  \bibinfo {pages} {079} (\bibinfo {year} {2017})},\ \Eprint
  {http://arxiv.org/abs/1710.03234} {arXiv:1710.03234 [hep-th]} \BibitemShut
  {NoStop}%
\bibitem [{\citenamefont {Blaizot}\ and\ \citenamefont
  {Yan}(2018)}]{Blaizot:2017ucy}%
  \BibitemOpen
  \bibfield  {author} {\bibinfo {author} {\bibfnamefont {Jean-Paul}\
  \bibnamefont {Blaizot}}\ and\ \bibinfo {author} {\bibfnamefont
  {Li}~\bibnamefont {Yan}},\ }\bibfield  {title} {\enquote {\bibinfo {title}
  {{Fluid dynamics of out of equilibrium boost invariant plasmas}},}\ }\href
  {\doibase 10.1016/j.physletb.2018.02.058} {\bibfield  {journal} {\bibinfo
  {journal} {Phys. Lett. B}\ }\textbf {\bibinfo {volume} {780}},\ \bibinfo
  {pages} {283--286} (\bibinfo {year} {2018})},\ \Eprint
  {http://arxiv.org/abs/1712.03856} {arXiv:1712.03856 [nucl-th]} \BibitemShut
  {NoStop}%
\bibitem [{\citenamefont {Spali\'nski}(2018)}]{Spalinski:2017mel}%
  \BibitemOpen
  \bibfield  {author} {\bibinfo {author} {\bibfnamefont {Micha\l{}}\
  \bibnamefont {Spali\'nski}},\ }\bibfield  {title} {\enquote {\bibinfo {title}
  {{On the hydrodynamic attractor of Yang\textendash{}Mills plasma}},}\ }\href
  {\doibase 10.1016/j.physletb.2017.11.059} {\bibfield  {journal} {\bibinfo
  {journal} {Phys. Lett. B}\ }\textbf {\bibinfo {volume} {776}},\ \bibinfo
  {pages} {468--472} (\bibinfo {year} {2018})},\ \Eprint
  {http://arxiv.org/abs/1708.01921} {arXiv:1708.01921 [hep-th]} \BibitemShut
  {NoStop}%
\bibitem [{\citenamefont {Ambrus}\ \emph {et~al.}(2021)\citenamefont {Ambrus},
  \citenamefont {Busuioc}, \citenamefont {Fotakis}, \citenamefont
  {Gallmeister},\ and\ \citenamefont {Greiner}}]{Ambrus:2021sjg}%
  \BibitemOpen
  \bibfield  {author} {\bibinfo {author} {\bibfnamefont {Victor~E.}\
  \bibnamefont {Ambrus}}, \bibinfo {author} {\bibfnamefont {Sergiu}\
  \bibnamefont {Busuioc}}, \bibinfo {author} {\bibfnamefont {Jan~A.}\
  \bibnamefont {Fotakis}}, \bibinfo {author} {\bibfnamefont {Kai}\ \bibnamefont
  {Gallmeister}}, \ and\ \bibinfo {author} {\bibfnamefont {Carsten}\
  \bibnamefont {Greiner}},\ }\bibfield  {title} {\enquote {\bibinfo {title}
  {{Bjorken flow attractors with transverse dynamics}},}\ }\href {\doibase
  10.1103/PhysRevD.104.094022} {\bibfield  {journal} {\bibinfo  {journal}
  {Phys. Rev. D}\ }\textbf {\bibinfo {volume} {104}},\ \bibinfo {pages}
  {094022} (\bibinfo {year} {2021})},\ \Eprint
  {http://arxiv.org/abs/2102.11785} {arXiv:2102.11785 [nucl-th]} \BibitemShut
  {NoStop}%
\bibitem [{\citenamefont {Kurkela}\ \emph {et~al.}(2020)\citenamefont
  {Kurkela}, \citenamefont {van~der Schee}, \citenamefont {Wiedemann},\ and\
  \citenamefont {Wu}}]{Kurkela:2019set}%
  \BibitemOpen
  \bibfield  {author} {\bibinfo {author} {\bibfnamefont {Aleksi}\ \bibnamefont
  {Kurkela}}, \bibinfo {author} {\bibfnamefont {Wilke}\ \bibnamefont {van~der
  Schee}}, \bibinfo {author} {\bibfnamefont {Urs~Achim}\ \bibnamefont
  {Wiedemann}}, \ and\ \bibinfo {author} {\bibfnamefont {Bin}\ \bibnamefont
  {Wu}},\ }\bibfield  {title} {\enquote {\bibinfo {title} {{Early- and
  Late-Time Behavior of Attractors in Heavy-Ion Collisions}},}\ }\href
  {\doibase 10.1103/PhysRevLett.124.102301} {\bibfield  {journal} {\bibinfo
  {journal} {Phys. Rev. Lett.}\ }\textbf {\bibinfo {volume} {124}},\ \bibinfo
  {pages} {102301} (\bibinfo {year} {2020})},\ \Eprint
  {http://arxiv.org/abs/1907.08101} {arXiv:1907.08101 [hep-ph]} \BibitemShut
  {NoStop}%
\bibitem [{\citenamefont {Almaalol}\ \emph {et~al.}(2020)\citenamefont
  {Almaalol}, \citenamefont {Kurkela},\ and\ \citenamefont
  {Strickland}}]{Almaalol:2020rnu}%
  \BibitemOpen
  \bibfield  {author} {\bibinfo {author} {\bibfnamefont {Dekrayat}\
  \bibnamefont {Almaalol}}, \bibinfo {author} {\bibfnamefont {Aleksi}\
  \bibnamefont {Kurkela}}, \ and\ \bibinfo {author} {\bibfnamefont {Michael}\
  \bibnamefont {Strickland}},\ }\bibfield  {title} {\enquote {\bibinfo {title}
  {{Nonequilibrium Attractor in High-Temperature QCD Plasmas}},}\ }\href
  {\doibase 10.1103/PhysRevLett.125.122302} {\bibfield  {journal} {\bibinfo
  {journal} {Phys. Rev. Lett.}\ }\textbf {\bibinfo {volume} {125}},\ \bibinfo
  {pages} {122302} (\bibinfo {year} {2020})},\ \Eprint
  {http://arxiv.org/abs/2004.05195} {arXiv:2004.05195 [hep-ph]} \BibitemShut
  {NoStop}%
\bibitem [{\citenamefont {Blaizot}\ and\ \citenamefont
  {Yan}(2020)}]{Blaizot:2019scw}%
  \BibitemOpen
  \bibfield  {author} {\bibinfo {author} {\bibfnamefont {Jean-Paul}\
  \bibnamefont {Blaizot}}\ and\ \bibinfo {author} {\bibfnamefont
  {Li}~\bibnamefont {Yan}},\ }\bibfield  {title} {\enquote {\bibinfo {title}
  {{Emergence of hydrodynamical behavior in expanding ultra-relativistic
  plasmas}},}\ }\href {\doibase 10.1016/j.aop.2019.167993} {\bibfield
  {journal} {\bibinfo  {journal} {Annals Phys.}\ }\textbf {\bibinfo {volume}
  {412}},\ \bibinfo {pages} {167993} (\bibinfo {year} {2020})},\ \Eprint
  {http://arxiv.org/abs/1904.08677} {arXiv:1904.08677 [nucl-th]} \BibitemShut
  {NoStop}%
\bibitem [{\citenamefont {Blaizot}\ and\ \citenamefont
  {Yan}(2021{\natexlab{a}})}]{Blaizot:2020gql}%
  \BibitemOpen
  \bibfield  {author} {\bibinfo {author} {\bibfnamefont {Jean-Paul}\
  \bibnamefont {Blaizot}}\ and\ \bibinfo {author} {\bibfnamefont
  {Li}~\bibnamefont {Yan}},\ }\bibfield  {title} {\enquote {\bibinfo {title}
  {{Analytical attractor for Bjorken flows}},}\ }\href {\doibase
  10.1016/j.physletb.2021.136478} {\bibfield  {journal} {\bibinfo  {journal}
  {Phys. Lett. B}\ }\textbf {\bibinfo {volume} {820}},\ \bibinfo {pages}
  {136478} (\bibinfo {year} {2021}{\natexlab{a}})},\ \Eprint
  {http://arxiv.org/abs/2006.08815} {arXiv:2006.08815 [nucl-th]} \BibitemShut
  {NoStop}%
\bibitem [{\citenamefont {Blaizot}\ and\ \citenamefont
  {Yan}(2021{\natexlab{b}})}]{Blaizot:2021cdv}%
  \BibitemOpen
  \bibfield  {author} {\bibinfo {author} {\bibfnamefont {Jean-Paul}\
  \bibnamefont {Blaizot}}\ and\ \bibinfo {author} {\bibfnamefont
  {Li}~\bibnamefont {Yan}},\ }\bibfield  {title} {\enquote {\bibinfo {title}
  {{Attractor and fixed points in Bjorken flows}},}\ }\href {\doibase
  10.1103/PhysRevC.104.055201} {\bibfield  {journal} {\bibinfo  {journal}
  {Phys. Rev. C}\ }\textbf {\bibinfo {volume} {104}},\ \bibinfo {pages}
  {055201} (\bibinfo {year} {2021}{\natexlab{b}})},\ \Eprint
  {http://arxiv.org/abs/2106.10508} {arXiv:2106.10508 [nucl-th]} \BibitemShut
  {NoStop}%
\bibitem [{\citenamefont {Chen}\ and\ \citenamefont
  {Shi}(2025)}]{Chen:2024pez}%
  \BibitemOpen
  \bibfield  {author} {\bibinfo {author} {\bibfnamefont {Shile}\ \bibnamefont
  {Chen}}\ and\ \bibinfo {author} {\bibfnamefont {Shuzhe}\ \bibnamefont
  {Shi}},\ }\bibfield  {title} {\enquote {\bibinfo {title} {{Attractor for
  (1+1)D viscous hydrodynamics with general rapidity distribution}},}\ }\href
  {\doibase 10.1103/PhysRevC.111.L021902} {\bibfield  {journal} {\bibinfo
  {journal} {Phys. Rev. C}\ }\textbf {\bibinfo {volume} {111}},\ \bibinfo
  {pages} {L021902} (\bibinfo {year} {2025})},\ \Eprint
  {http://arxiv.org/abs/2407.15209} {arXiv:2407.15209 [hep-ph]} \BibitemShut
  {NoStop}%
\bibitem [{\citenamefont {Baier}\ \emph {et~al.}(2001)\citenamefont {Baier},
  \citenamefont {Mueller}, \citenamefont {Schiff},\ and\ \citenamefont
  {Son}}]{Baier:2000sb}%
  \BibitemOpen
  \bibfield  {author} {\bibinfo {author} {\bibfnamefont {R.}~\bibnamefont
  {Baier}}, \bibinfo {author} {\bibfnamefont {Alfred~H.}\ \bibnamefont
  {Mueller}}, \bibinfo {author} {\bibfnamefont {D.}~\bibnamefont {Schiff}}, \
  and\ \bibinfo {author} {\bibfnamefont {D.~T.}\ \bibnamefont {Son}},\
  }\bibfield  {title} {\enquote {\bibinfo {title} {{'Bottom up' thermalization
  in heavy ion collisions}},}\ }\href {\doibase 10.1016/S0370-2693(01)00191-5}
  {\bibfield  {journal} {\bibinfo  {journal} {Phys. Lett. B}\ }\textbf
  {\bibinfo {volume} {502}},\ \bibinfo {pages} {51--58} (\bibinfo {year}
  {2001})},\ \Eprint {http://arxiv.org/abs/hep-ph/0009237}
  {arXiv:hep-ph/0009237} \BibitemShut {NoStop}%
\bibitem [{\citenamefont {Kurkela}\ and\ \citenamefont
  {Zhu}(2015)}]{Kurkela:2015qoa}%
  \BibitemOpen
  \bibfield  {author} {\bibinfo {author} {\bibfnamefont {Aleksi}\ \bibnamefont
  {Kurkela}}\ and\ \bibinfo {author} {\bibfnamefont {Yan}\ \bibnamefont
  {Zhu}},\ }\bibfield  {title} {\enquote {\bibinfo {title} {{Isotropization and
  hydrodynamization in weakly coupled heavy-ion collisions}},}\ }\href
  {\doibase 10.1103/PhysRevLett.115.182301} {\bibfield  {journal} {\bibinfo
  {journal} {Phys. Rev. Lett.}\ }\textbf {\bibinfo {volume} {115}},\ \bibinfo
  {pages} {182301} (\bibinfo {year} {2015})},\ \Eprint
  {http://arxiv.org/abs/1506.06647} {arXiv:1506.06647 [hep-ph]} \BibitemShut
  {NoStop}%
\bibitem [{\citenamefont {Blaizot}\ \emph {et~al.}(2012)\citenamefont
  {Blaizot}, \citenamefont {Gelis}, \citenamefont {Liao}, \citenamefont
  {McLerran},\ and\ \citenamefont {Venugopalan}}]{Blaizot:2011xf}%
  \BibitemOpen
  \bibfield  {author} {\bibinfo {author} {\bibfnamefont {Jean-Paul}\
  \bibnamefont {Blaizot}}, \bibinfo {author} {\bibfnamefont {Francois}\
  \bibnamefont {Gelis}}, \bibinfo {author} {\bibfnamefont {Jin-Feng}\
  \bibnamefont {Liao}}, \bibinfo {author} {\bibfnamefont {Larry}\ \bibnamefont
  {McLerran}}, \ and\ \bibinfo {author} {\bibfnamefont {Raju}\ \bibnamefont
  {Venugopalan}},\ }\bibfield  {title} {\enquote {\bibinfo {title}
  {{Bose--Einstein Condensation and Thermalization of the Quark Gluon
  Plasma}},}\ }\href {\doibase 10.1016/j.nuclphysa.2011.10.005} {\bibfield
  {journal} {\bibinfo  {journal} {Nucl. Phys. A}\ }\textbf {\bibinfo {volume}
  {873}},\ \bibinfo {pages} {68--80} (\bibinfo {year} {2012})},\ \Eprint
  {http://arxiv.org/abs/1107.5296} {arXiv:1107.5296 [hep-ph]} \BibitemShut
  {NoStop}%
\bibitem [{\citenamefont {Blaizot}\ \emph {et~al.}(2013)\citenamefont
  {Blaizot}, \citenamefont {Liao},\ and\ \citenamefont
  {McLerran}}]{Blaizot:2013lga}%
  \BibitemOpen
  \bibfield  {author} {\bibinfo {author} {\bibfnamefont {Jean-Paul}\
  \bibnamefont {Blaizot}}, \bibinfo {author} {\bibfnamefont {Jinfeng}\
  \bibnamefont {Liao}}, \ and\ \bibinfo {author} {\bibfnamefont {Larry}\
  \bibnamefont {McLerran}},\ }\bibfield  {title} {\enquote {\bibinfo {title}
  {{Gluon Transport Equation in the Small Angle Approximation and the Onset of
  Bose-Einstein Condensation}},}\ }\href {\doibase
  10.1016/j.nuclphysa.2013.10.010} {\bibfield  {journal} {\bibinfo  {journal}
  {Nucl. Phys. A}\ }\textbf {\bibinfo {volume} {920}},\ \bibinfo {pages}
  {58--77} (\bibinfo {year} {2013})},\ \Eprint {http://arxiv.org/abs/1305.2119}
  {arXiv:1305.2119 [hep-ph]} \BibitemShut {NoStop}%
\bibitem [{\citenamefont {Berges}\ and\ \citenamefont
  {Sexty}(2012)}]{Berges:2012us}%
  \BibitemOpen
  \bibfield  {author} {\bibinfo {author} {\bibfnamefont {J.}~\bibnamefont
  {Berges}}\ and\ \bibinfo {author} {\bibfnamefont {D.}~\bibnamefont {Sexty}},\
  }\bibfield  {title} {\enquote {\bibinfo {title} {{Bose condensation far from
  equilibrium}},}\ }\href {\doibase 10.1103/PhysRevLett.108.161601} {\bibfield
  {journal} {\bibinfo  {journal} {Phys. Rev. Lett.}\ }\textbf {\bibinfo
  {volume} {108}},\ \bibinfo {pages} {161601} (\bibinfo {year} {2012})},\
  \Eprint {http://arxiv.org/abs/1201.0687} {arXiv:1201.0687 [hep-ph]}
  \BibitemShut {NoStop}%
\bibitem [{\citenamefont {Berges}\ \emph {et~al.}(2015)\citenamefont {Berges},
  \citenamefont {Boguslavski}, \citenamefont {Schlichting},\ and\ \citenamefont
  {Venugopalan}}]{Berges:2014bba}%
  \BibitemOpen
  \bibfield  {author} {\bibinfo {author} {\bibfnamefont {J.}~\bibnamefont
  {Berges}}, \bibinfo {author} {\bibfnamefont {K.}~\bibnamefont {Boguslavski}},
  \bibinfo {author} {\bibfnamefont {S.}~\bibnamefont {Schlichting}}, \ and\
  \bibinfo {author} {\bibfnamefont {R.}~\bibnamefont {Venugopalan}},\
  }\bibfield  {title} {\enquote {\bibinfo {title} {{Universality far from
  equilibrium: From superfluid Bose gases to heavy-ion collisions}},}\ }\href
  {\doibase 10.1103/PhysRevLett.114.061601} {\bibfield  {journal} {\bibinfo
  {journal} {Phys. Rev. Lett.}\ }\textbf {\bibinfo {volume} {114}},\ \bibinfo
  {pages} {061601} (\bibinfo {year} {2015})},\ \Eprint
  {http://arxiv.org/abs/1408.1670} {arXiv:1408.1670 [hep-ph]} \BibitemShut
  {NoStop}%
\bibitem [{\citenamefont {Blaizot}\ \emph {et~al.}(2014)\citenamefont
  {Blaizot}, \citenamefont {Wu},\ and\ \citenamefont {Yan}}]{Blaizot:2014jna}%
  \BibitemOpen
  \bibfield  {author} {\bibinfo {author} {\bibfnamefont {Jean-Paul}\
  \bibnamefont {Blaizot}}, \bibinfo {author} {\bibfnamefont {Bin}\ \bibnamefont
  {Wu}}, \ and\ \bibinfo {author} {\bibfnamefont {Li}~\bibnamefont {Yan}},\
  }\bibfield  {title} {\enquote {\bibinfo {title} {{Quark production,
  Bose\textendash{}Einstein condensates and thermalization of the
  quark\textendash{}gluon plasma}},}\ }\href {\doibase
  10.1016/j.nuclphysa.2014.07.041} {\bibfield  {journal} {\bibinfo  {journal}
  {Nucl. Phys. A}\ }\textbf {\bibinfo {volume} {930}},\ \bibinfo {pages}
  {139--162} (\bibinfo {year} {2014})},\ \Eprint
  {http://arxiv.org/abs/1402.5049} {arXiv:1402.5049 [hep-ph]} \BibitemShut
  {NoStop}%
\bibitem [{\citenamefont {Xu}\ \emph {et~al.}(2015)\citenamefont {Xu},
  \citenamefont {Zhou}, \citenamefont {Zhuang},\ and\ \citenamefont
  {Greiner}}]{Xu:2014ega}%
  \BibitemOpen
  \bibfield  {author} {\bibinfo {author} {\bibfnamefont {Zhe}\ \bibnamefont
  {Xu}}, \bibinfo {author} {\bibfnamefont {Kai}\ \bibnamefont {Zhou}}, \bibinfo
  {author} {\bibfnamefont {Pengfei}\ \bibnamefont {Zhuang}}, \ and\ \bibinfo
  {author} {\bibfnamefont {Carsten}\ \bibnamefont {Greiner}},\ }\bibfield
  {title} {\enquote {\bibinfo {title} {{Thermalization of gluons with
  Bose-Einstein condensation}},}\ }\href {\doibase
  10.1103/PhysRevLett.114.182301} {\bibfield  {journal} {\bibinfo  {journal}
  {Phys. Rev. Lett.}\ }\textbf {\bibinfo {volume} {114}},\ \bibinfo {pages}
  {182301} (\bibinfo {year} {2015})},\ \Eprint {http://arxiv.org/abs/1410.5616}
  {arXiv:1410.5616 [hep-ph]} \BibitemShut {NoStop}%
\bibitem [{\citenamefont {Martinez}\ and\ \citenamefont
  {Strickland}(2010)}]{Martinez:2010sc}%
  \BibitemOpen
  \bibfield  {author} {\bibinfo {author} {\bibfnamefont {Mauricio}\
  \bibnamefont {Martinez}}\ and\ \bibinfo {author} {\bibfnamefont {Michael}\
  \bibnamefont {Strickland}},\ }\bibfield  {title} {\enquote {\bibinfo {title}
  {{Dissipative Dynamics of Highly Anisotropic Systems}},}\ }\href {\doibase
  10.1016/j.nuclphysa.2010.08.011} {\bibfield  {journal} {\bibinfo  {journal}
  {Nucl. Phys. A}\ }\textbf {\bibinfo {volume} {848}},\ \bibinfo {pages}
  {183--197} (\bibinfo {year} {2010})},\ \Eprint
  {http://arxiv.org/abs/1007.0889} {arXiv:1007.0889 [nucl-th]} \BibitemShut
  {NoStop}%
\bibitem [{\citenamefont {Martinez}\ \emph {et~al.}(2012)\citenamefont
  {Martinez}, \citenamefont {Ryblewski},\ and\ \citenamefont
  {Strickland}}]{Martinez:2012tu}%
  \BibitemOpen
  \bibfield  {author} {\bibinfo {author} {\bibfnamefont {Mauricio}\
  \bibnamefont {Martinez}}, \bibinfo {author} {\bibfnamefont {Radoslaw}\
  \bibnamefont {Ryblewski}}, \ and\ \bibinfo {author} {\bibfnamefont {Michael}\
  \bibnamefont {Strickland}},\ }\bibfield  {title} {\enquote {\bibinfo {title}
  {{Boost-Invariant (2+1)-dimensional Anisotropic Hydrodynamics}},}\ }\href
  {\doibase 10.1103/PhysRevC.85.064913} {\bibfield  {journal} {\bibinfo
  {journal} {Phys. Rev. C}\ }\textbf {\bibinfo {volume} {85}},\ \bibinfo
  {pages} {064913} (\bibinfo {year} {2012})},\ \Eprint
  {http://arxiv.org/abs/1204.1473} {arXiv:1204.1473 [nucl-th]} \BibitemShut
  {NoStop}%
\bibitem [{\citenamefont {Ryblewski}\ and\ \citenamefont
  {Florkowski}(2012)}]{Ryblewski:2012rr}%
  \BibitemOpen
  \bibfield  {author} {\bibinfo {author} {\bibfnamefont {Radoslaw}\
  \bibnamefont {Ryblewski}}\ and\ \bibinfo {author} {\bibfnamefont {Wojciech}\
  \bibnamefont {Florkowski}},\ }\bibfield  {title} {\enquote {\bibinfo {title}
  {{Highly-anisotropic hydrodynamics in 3+1 space-time dimensions}},}\ }\href
  {\doibase 10.1103/PhysRevC.85.064901} {\bibfield  {journal} {\bibinfo
  {journal} {Phys. Rev. C}\ }\textbf {\bibinfo {volume} {85}},\ \bibinfo
  {pages} {064901} (\bibinfo {year} {2012})},\ \Eprint
  {http://arxiv.org/abs/1204.2624} {arXiv:1204.2624 [nucl-th]} \BibitemShut
  {NoStop}%
\bibitem [{\citenamefont {Strickland}\ \emph {et~al.}(2018)\citenamefont
  {Strickland}, \citenamefont {Noronha},\ and\ \citenamefont
  {Denicol}}]{Strickland:2017kux}%
  \BibitemOpen
  \bibfield  {author} {\bibinfo {author} {\bibfnamefont {Michael}\ \bibnamefont
  {Strickland}}, \bibinfo {author} {\bibfnamefont {Jorge}\ \bibnamefont
  {Noronha}}, \ and\ \bibinfo {author} {\bibfnamefont {Gabriel}\ \bibnamefont
  {Denicol}},\ }\bibfield  {title} {\enquote {\bibinfo {title} {{Anisotropic
  nonequilibrium hydrodynamic attractor}},}\ }\href {\doibase
  10.1103/PhysRevD.97.036020} {\bibfield  {journal} {\bibinfo  {journal} {Phys.
  Rev. D}\ }\textbf {\bibinfo {volume} {97}},\ \bibinfo {pages} {036020}
  (\bibinfo {year} {2018})},\ \Eprint {http://arxiv.org/abs/1709.06644}
  {arXiv:1709.06644 [nucl-th]} \BibitemShut {NoStop}%
\bibitem [{\citenamefont {Strickland}(2018)}]{Strickland:2018ayk}%
  \BibitemOpen
  \bibfield  {author} {\bibinfo {author} {\bibfnamefont {M.}~\bibnamefont
  {Strickland}},\ }\bibfield  {title} {\enquote {\bibinfo {title} {{The
  non-equilibrium attractor for kinetic theory in relaxation time
  approximation}},}\ }\href {\doibase 10.1007/JHEP12(2018)128} {\bibfield
  {journal} {\bibinfo  {journal} {JHEP}\ }\textbf {\bibinfo {volume} {12}},\
  \bibinfo {pages} {128} (\bibinfo {year} {2018})},\ \Eprint
  {http://arxiv.org/abs/1809.01200} {arXiv:1809.01200 [nucl-th]} \BibitemShut
  {NoStop}%
\bibitem [{\citenamefont {Romatschke}\ and\ \citenamefont
  {Romatschke}(2019)}]{Romatschke:2017ejr}%
  \BibitemOpen
  \bibfield  {author} {\bibinfo {author} {\bibfnamefont {Paul}\ \bibnamefont
  {Romatschke}}\ and\ \bibinfo {author} {\bibfnamefont {Ulrike}\ \bibnamefont
  {Romatschke}},\ }\href {\doibase 10.1017/9781108651998} {\emph {\bibinfo
  {title} {{Relativistic Fluid Dynamics In and Out of Equilibrium}}}},\
  Cambridge Monographs on Mathematical Physics\ (\bibinfo  {publisher}
  {Cambridge University Press},\ \bibinfo {year} {2019})\ \Eprint
  {http://arxiv.org/abs/1712.05815} {arXiv:1712.05815 [nucl-th]} \BibitemShut
  {NoStop}%
\bibitem [{\citenamefont {Berges}\ \emph {et~al.}(2021)\citenamefont {Berges},
  \citenamefont {Heller}, \citenamefont {Mazeliauskas},\ and\ \citenamefont
  {Venugopalan}}]{Berges:2020fwq}%
  \BibitemOpen
  \bibfield  {author} {\bibinfo {author} {\bibfnamefont {J\"urgen}\
  \bibnamefont {Berges}}, \bibinfo {author} {\bibfnamefont {Michal~P.}\
  \bibnamefont {Heller}}, \bibinfo {author} {\bibfnamefont {Aleksas}\
  \bibnamefont {Mazeliauskas}}, \ and\ \bibinfo {author} {\bibfnamefont {Raju}\
  \bibnamefont {Venugopalan}},\ }\bibfield  {title} {\enquote {\bibinfo {title}
  {{QCD thermalization: Ab initio approaches and interdisciplinary
  connections}},}\ }\href {\doibase 10.1103/RevModPhys.93.035003} {\bibfield
  {journal} {\bibinfo  {journal} {Rev. Mod. Phys.}\ }\textbf {\bibinfo {volume}
  {93}},\ \bibinfo {pages} {035003} (\bibinfo {year} {2021})},\ \Eprint
  {http://arxiv.org/abs/2005.12299} {arXiv:2005.12299 [hep-th]} \BibitemShut
  {NoStop}%
\bibitem [{\citenamefont {Lu}\ and\ \citenamefont {Shi}(2025)}]{Lu:2025yry}%
  \BibitemOpen
  \bibfield  {author} {\bibinfo {author} {\bibfnamefont {Xingjian}\
  \bibnamefont {Lu}}\ and\ \bibinfo {author} {\bibfnamefont {Shuzhe}\
  \bibnamefont {Shi}},\ }\bibfield  {title} {\enquote {\bibinfo {title}
  {{Spectral BBGKY: a scalable scheme for nonlinear Boltzmann and correlation
  kinetics}},}\ }\href@noop {} {\  (\bibinfo {year} {2025})},\ \Eprint
  {http://arxiv.org/abs/2507.14243} {arXiv:2507.14243 [nucl-th]} \BibitemShut
  {NoStop}%
\bibitem [{\citenamefont {Alqahtani}\ \emph {et~al.}(2018)\citenamefont
  {Alqahtani}, \citenamefont {Nopoush},\ and\ \citenamefont
  {Strickland}}]{Alqahtani:2017mhy}%
  \BibitemOpen
  \bibfield  {author} {\bibinfo {author} {\bibfnamefont {Mubarak}\ \bibnamefont
  {Alqahtani}}, \bibinfo {author} {\bibfnamefont {Mohammad}\ \bibnamefont
  {Nopoush}}, \ and\ \bibinfo {author} {\bibfnamefont {Michael}\ \bibnamefont
  {Strickland}},\ }\bibfield  {title} {\enquote {\bibinfo {title}
  {{Relativistic anisotropic hydrodynamics}},}\ }\href {\doibase
  10.1016/j.ppnp.2018.05.004} {\bibfield  {journal} {\bibinfo  {journal} {Prog.
  Part. Nucl. Phys.}\ }\textbf {\bibinfo {volume} {101}},\ \bibinfo {pages}
  {204--248} (\bibinfo {year} {2018})},\ \Eprint
  {http://arxiv.org/abs/1712.03282} {arXiv:1712.03282 [nucl-th]} \BibitemShut
  {NoStop}%
\bibitem [{\citenamefont {Florkowski}\ \emph {et~al.}(2018)\citenamefont
  {Florkowski}, \citenamefont {Heller},\ and\ \citenamefont
  {Spalinski}}]{Florkowski:2017olj}%
  \BibitemOpen
  \bibfield  {author} {\bibinfo {author} {\bibfnamefont {Wojciech}\
  \bibnamefont {Florkowski}}, \bibinfo {author} {\bibfnamefont {Michal~P.}\
  \bibnamefont {Heller}}, \ and\ \bibinfo {author} {\bibfnamefont {Michal}\
  \bibnamefont {Spalinski}},\ }\bibfield  {title} {\enquote {\bibinfo {title}
  {{New theories of relativistic hydrodynamics in the LHC era}},}\ }\href
  {\doibase 10.1088/1361-6633/aaa091} {\bibfield  {journal} {\bibinfo
  {journal} {Rept. Prog. Phys.}\ }\textbf {\bibinfo {volume} {81}},\ \bibinfo
  {pages} {046001} (\bibinfo {year} {2018})},\ \Eprint
  {http://arxiv.org/abs/1707.02282} {arXiv:1707.02282 [hep-ph]} \BibitemShut
  {NoStop}%
\bibitem [{\citenamefont {Shen}\ and\ \citenamefont
  {Yan}(2020)}]{Shen:2020mgh}%
  \BibitemOpen
  \bibfield  {author} {\bibinfo {author} {\bibfnamefont {Chun}\ \bibnamefont
  {Shen}}\ and\ \bibinfo {author} {\bibfnamefont {Li}~\bibnamefont {Yan}},\
  }\bibfield  {title} {\enquote {\bibinfo {title} {{Recent development of
  hydrodynamic modeling in heavy-ion collisions}},}\ }\href {\doibase
  10.1007/s41365-020-00829-z} {\bibfield  {journal} {\bibinfo  {journal} {Nucl.
  Sci. Tech.}\ }\textbf {\bibinfo {volume} {31}},\ \bibinfo {pages} {122}
  (\bibinfo {year} {2020})},\ \Eprint {http://arxiv.org/abs/2010.12377}
  {arXiv:2010.12377 [nucl-th]} \BibitemShut {NoStop}%
\bibitem [{\citenamefont {Soloviev}(2022)}]{Soloviev:2021lhs}%
  \BibitemOpen
  \bibfield  {author} {\bibinfo {author} {\bibfnamefont {Alexander}\
  \bibnamefont {Soloviev}},\ }\bibfield  {title} {\enquote {\bibinfo {title}
  {{Hydrodynamic attractors in heavy ion collisions: a review}},}\ }\href
  {\doibase 10.1140/epjc/s10052-022-10282-4} {\bibfield  {journal} {\bibinfo
  {journal} {Eur. Phys. J. C}\ }\textbf {\bibinfo {volume} {82}},\ \bibinfo
  {pages} {319} (\bibinfo {year} {2022})},\ \Eprint
  {http://arxiv.org/abs/2109.15081} {arXiv:2109.15081 [hep-th]} \BibitemShut
  {NoStop}%
\bibitem [{\citenamefont {Jankowski}\ and\ \citenamefont
  {Spali\'nski}(2023)}]{Jankowski:2023fdz}%
  \BibitemOpen
  \bibfield  {author} {\bibinfo {author} {\bibfnamefont {Jakub}\ \bibnamefont
  {Jankowski}}\ and\ \bibinfo {author} {\bibfnamefont {Micha\l{}}\ \bibnamefont
  {Spali\'nski}},\ }\bibfield  {title} {\enquote {\bibinfo {title}
  {{Hydrodynamic attractors in ultrarelativistic nuclear collisions}},}\ }\href
  {\doibase 10.1016/j.ppnp.2023.104048} {\bibfield  {journal} {\bibinfo
  {journal} {Prog. Part. Nucl. Phys.}\ }\textbf {\bibinfo {volume} {132}},\
  \bibinfo {pages} {104048} (\bibinfo {year} {2023})},\ \Eprint
  {http://arxiv.org/abs/2303.09414} {arXiv:2303.09414 [nucl-th]} \BibitemShut
  {NoStop}%
\bibitem [{\citenamefont {Strickland}(2024)}]{Strickland:2024moq}%
  \BibitemOpen
  \bibfield  {author} {\bibinfo {author} {\bibfnamefont {Michael}\ \bibnamefont
  {Strickland}},\ }\bibfield  {title} {\enquote {\bibinfo {title}
  {{Hydrodynamization and resummed viscous hydrodynamics}},}\ }\href {\doibase
  10.1142/9789811294679_0003} {\bibfield  {journal} {\bibinfo  {journal} {Int.
  J. Mod. Phys. E}\ }\textbf {\bibinfo {volume} {33}},\ \bibinfo {pages}
  {2430004} (\bibinfo {year} {2024})},\ \Eprint
  {http://arxiv.org/abs/2402.09571} {arXiv:2402.09571 [nucl-th]} \BibitemShut
  {NoStop}%
\bibitem [{\citenamefont {Schwinger}(1962)}]{Schwinger:1962tp}%
  \BibitemOpen
  \bibfield  {author} {\bibinfo {author} {\bibfnamefont {Julian~S.}\
  \bibnamefont {Schwinger}},\ }\bibfield  {title} {\enquote {\bibinfo {title}
  {{Gauge Invariance and Mass. 2.}}}\ }\href {\doibase
  10.1103/PhysRev.128.2425} {\bibfield  {journal} {\bibinfo  {journal} {Phys.
  Rev.}\ }\textbf {\bibinfo {volume} {128}},\ \bibinfo {pages} {2425--2429}
  (\bibinfo {year} {1962})}\BibitemShut {NoStop}%
\bibitem [{\citenamefont {Lowenstein}\ and\ \citenamefont
  {Swieca}(1971)}]{Lowenstein:1971fc}%
  \BibitemOpen
  \bibfield  {author} {\bibinfo {author} {\bibfnamefont {J.~H.}\ \bibnamefont
  {Lowenstein}}\ and\ \bibinfo {author} {\bibfnamefont {J.~A.}\ \bibnamefont
  {Swieca}},\ }\bibfield  {title} {\enquote {\bibinfo {title} {{Quantum
  electrodynamics in two-dimensions}},}\ }\href {\doibase
  10.1016/0003-4916(71)90246-6} {\bibfield  {journal} {\bibinfo  {journal}
  {Annals Phys.}\ }\textbf {\bibinfo {volume} {68}},\ \bibinfo {pages}
  {172--195} (\bibinfo {year} {1971})}\BibitemShut {NoStop}%
\bibitem [{\citenamefont {Jayewardena}(1988)}]{Jayewardena:1988td}%
  \BibitemOpen
  \bibfield  {author} {\bibinfo {author} {\bibfnamefont {C.}~\bibnamefont
  {Jayewardena}},\ }\bibfield  {title} {\enquote {\bibinfo {title} {{SCHWINGER
  MODEL ON S(2)}},}\ }\href@noop {} {\bibfield  {journal} {\bibinfo  {journal}
  {Helv. Phys. Acta}\ }\textbf {\bibinfo {volume} {61}},\ \bibinfo {pages}
  {636--711} (\bibinfo {year} {1988})}\BibitemShut {NoStop}%
\bibitem [{\citenamefont {Smilga}(1992)}]{Smilga:1992hx}%
  \BibitemOpen
  \bibfield  {author} {\bibinfo {author} {\bibfnamefont {Andrei~V.}\
  \bibnamefont {Smilga}},\ }\bibfield  {title} {\enquote {\bibinfo {title} {{On
  the fermion condensate in Schwinger model}},}\ }\href {\doibase
  10.1016/0370-2693(92)90209-M} {\bibfield  {journal} {\bibinfo  {journal}
  {Phys. Lett. B}\ }\textbf {\bibinfo {volume} {278}},\ \bibinfo {pages}
  {371--376} (\bibinfo {year} {1992})}\BibitemShut {NoStop}%
\bibitem [{\citenamefont {Adam}(1994)}]{Adam:1993fc}%
  \BibitemOpen
  \bibfield  {author} {\bibinfo {author} {\bibfnamefont {C.}~\bibnamefont
  {Adam}},\ }\bibfield  {title} {\enquote {\bibinfo {title} {{Instantons and
  vacuum expectation values in the Schwinger model}},}\ }\href {\doibase
  10.1007/BF01577557} {\bibfield  {journal} {\bibinfo  {journal} {Z. Phys. C}\
  }\textbf {\bibinfo {volume} {63}},\ \bibinfo {pages} {169--180} (\bibinfo
  {year} {1994})}\BibitemShut {NoStop}%
\bibitem [{\citenamefont {Adam}(1997{\natexlab{a}})}]{Adam:1997wt}%
  \BibitemOpen
  \bibfield  {author} {\bibinfo {author} {\bibfnamefont {Christoph}\
  \bibnamefont {Adam}},\ }\bibfield  {title} {\enquote {\bibinfo {title}
  {{Massive Schwinger model within mass perturbation theory}},}\ }\href
  {\doibase 10.1006/aphy.1997.5697} {\bibfield  {journal} {\bibinfo  {journal}
  {Annals Phys.}\ }\textbf {\bibinfo {volume} {259}},\ \bibinfo {pages} {1--63}
  (\bibinfo {year} {1997}{\natexlab{a}})},\ \Eprint
  {http://arxiv.org/abs/hep-th/9704064} {arXiv:hep-th/9704064} \BibitemShut
  {NoStop}%
\bibitem [{\citenamefont {Coleman}(1976)}]{Coleman:1976uz}%
  \BibitemOpen
  \bibfield  {author} {\bibinfo {author} {\bibfnamefont {Sidney~R.}\
  \bibnamefont {Coleman}},\ }\bibfield  {title} {\enquote {\bibinfo {title}
  {{More About the Massive Schwinger Model}},}\ }\href {\doibase
  10.1016/0003-4916(76)90280-3} {\bibfield  {journal} {\bibinfo  {journal}
  {Annals Phys.}\ }\textbf {\bibinfo {volume} {101}},\ \bibinfo {pages} {239}
  (\bibinfo {year} {1976})}\BibitemShut {NoStop}%
\bibitem [{\citenamefont {Adam}(1996{\natexlab{a}})}]{Adam:1995us}%
  \BibitemOpen
  \bibfield  {author} {\bibinfo {author} {\bibfnamefont {Christoph}\
  \bibnamefont {Adam}},\ }\bibfield  {title} {\enquote {\bibinfo {title} {{The
  Schwinger mass in the massive Schwinger model}},}\ }\href {\doibase
  10.1016/0370-2693(96)00695-8} {\bibfield  {journal} {\bibinfo  {journal}
  {Phys. Lett. B}\ }\textbf {\bibinfo {volume} {382}},\ \bibinfo {pages}
  {383--388} (\bibinfo {year} {1996}{\natexlab{a}})},\ \Eprint
  {http://arxiv.org/abs/hep-ph/9507331} {arXiv:hep-ph/9507331} \BibitemShut
  {NoStop}%
\bibitem [{\citenamefont {Adam}(1996{\natexlab{b}})}]{Adam:1996np}%
  \BibitemOpen
  \bibfield  {author} {\bibinfo {author} {\bibfnamefont {C.}~\bibnamefont
  {Adam}},\ }\bibfield  {title} {\enquote {\bibinfo {title} {{General bound
  state structure of the massive Schwinger model}},}\ }\href {\doibase
  10.1016/0370-2693(96)00639-9} {\bibfield  {journal} {\bibinfo  {journal}
  {Phys. Lett. B}\ }\textbf {\bibinfo {volume} {382}},\ \bibinfo {pages}
  {111--116} (\bibinfo {year} {1996}{\natexlab{b}})},\ \Eprint
  {http://arxiv.org/abs/hep-th/9602175} {arXiv:hep-th/9602175} \BibitemShut
  {NoStop}%
\bibitem [{\citenamefont {Adam}(1997{\natexlab{b}})}]{Adam:1996qk}%
  \BibitemOpen
  \bibfield  {author} {\bibinfo {author} {\bibfnamefont {Christoph}\
  \bibnamefont {Adam}},\ }\bibfield  {title} {\enquote {\bibinfo {title} {{The
  boson-boson bound state in the massive Schwinger model}},}\ }\href {\doibase
  10.1007/s002880050438} {\bibfield  {journal} {\bibinfo  {journal} {Z. Phys.
  C}\ }\textbf {\bibinfo {volume} {74}},\ \bibinfo {pages} {727--730} (\bibinfo
  {year} {1997}{\natexlab{b}})},\ \Eprint {http://arxiv.org/abs/hep-ph/9601227}
  {arXiv:hep-ph/9601227} \BibitemShut {NoStop}%
\bibitem [{\citenamefont {Klco}\ \emph {et~al.}(2018)\citenamefont {Klco},
  \citenamefont {Dumitrescu}, \citenamefont {McCaskey}, \citenamefont {Morris},
  \citenamefont {Pooser}, \citenamefont {Sanz}, \citenamefont {Solano},
  \citenamefont {Lougovski},\ and\ \citenamefont {Savage}}]{Klco:2018kyo}%
  \BibitemOpen
  \bibfield  {author} {\bibinfo {author} {\bibfnamefont {N.}~\bibnamefont
  {Klco}}, \bibinfo {author} {\bibfnamefont {E.~F.}\ \bibnamefont
  {Dumitrescu}}, \bibinfo {author} {\bibfnamefont {A.~J.}\ \bibnamefont
  {McCaskey}}, \bibinfo {author} {\bibfnamefont {T.~D.}\ \bibnamefont
  {Morris}}, \bibinfo {author} {\bibfnamefont {R.~C.}\ \bibnamefont {Pooser}},
  \bibinfo {author} {\bibfnamefont {M.}~\bibnamefont {Sanz}}, \bibinfo {author}
  {\bibfnamefont {E.}~\bibnamefont {Solano}}, \bibinfo {author} {\bibfnamefont
  {P.}~\bibnamefont {Lougovski}}, \ and\ \bibinfo {author} {\bibfnamefont
  {M.~J.}\ \bibnamefont {Savage}},\ }\bibfield  {title} {\enquote {\bibinfo
  {title} {{Quantum-classical computation of Schwinger model dynamics using
  quantum computers}},}\ }\href {\doibase 10.1103/PhysRevA.98.032331}
  {\bibfield  {journal} {\bibinfo  {journal} {Phys. Rev. A}\ }\textbf {\bibinfo
  {volume} {98}},\ \bibinfo {pages} {032331} (\bibinfo {year} {2018})},\
  \Eprint {http://arxiv.org/abs/1803.03326} {arXiv:1803.03326 [quant-ph]}
  \BibitemShut {NoStop}%
\bibitem [{\citenamefont {Farrell}\ \emph
  {et~al.}(2024{\natexlab{a}})\citenamefont {Farrell}, \citenamefont {Illa},
  \citenamefont {Ciavarella},\ and\ \citenamefont {Savage}}]{Farrell:2023fgd}%
  \BibitemOpen
  \bibfield  {author} {\bibinfo {author} {\bibfnamefont {Roland~C.}\
  \bibnamefont {Farrell}}, \bibinfo {author} {\bibfnamefont {Marc}\
  \bibnamefont {Illa}}, \bibinfo {author} {\bibfnamefont {Anthony~N.}\
  \bibnamefont {Ciavarella}}, \ and\ \bibinfo {author} {\bibfnamefont
  {Martin~J.}\ \bibnamefont {Savage}},\ }\bibfield  {title} {\enquote {\bibinfo
  {title} {{Scalable Circuits for Preparing Ground States on Digital Quantum
  Computers: The Schwinger Model Vacuum on 100 Qubits}},}\ }\href {\doibase
  10.1103/PRXQuantum.5.020315} {\bibfield  {journal} {\bibinfo  {journal} {PRX
  Quantum}\ }\textbf {\bibinfo {volume} {5}},\ \bibinfo {pages} {020315}
  (\bibinfo {year} {2024}{\natexlab{a}})},\ \Eprint
  {http://arxiv.org/abs/2308.04481} {arXiv:2308.04481 [quant-ph]} \BibitemShut
  {NoStop}%
\bibitem [{\citenamefont {Farrell}\ \emph
  {et~al.}(2024{\natexlab{b}})\citenamefont {Farrell}, \citenamefont {Illa},
  \citenamefont {Ciavarella},\ and\ \citenamefont {Savage}}]{Farrell:2024fit}%
  \BibitemOpen
  \bibfield  {author} {\bibinfo {author} {\bibfnamefont {Roland~C.}\
  \bibnamefont {Farrell}}, \bibinfo {author} {\bibfnamefont {Marc}\
  \bibnamefont {Illa}}, \bibinfo {author} {\bibfnamefont {Anthony~N.}\
  \bibnamefont {Ciavarella}}, \ and\ \bibinfo {author} {\bibfnamefont
  {Martin~J.}\ \bibnamefont {Savage}},\ }\bibfield  {title} {\enquote {\bibinfo
  {title} {{Quantum simulations of hadron dynamics in the Schwinger model using
  112 qubits}},}\ }\href {\doibase 10.1103/PhysRevD.109.114510} {\bibfield
  {journal} {\bibinfo  {journal} {Phys. Rev. D}\ }\textbf {\bibinfo {volume}
  {109}},\ \bibinfo {pages} {114510} (\bibinfo {year} {2024}{\natexlab{b}})},\
  \Eprint {http://arxiv.org/abs/2401.08044} {arXiv:2401.08044 [quant-ph]}
  \BibitemShut {NoStop}%
\bibitem [{\citenamefont {Zache}\ \emph {et~al.}(2019)\citenamefont {Zache},
  \citenamefont {Mueller}, \citenamefont {Schneider}, \citenamefont
  {Jendrzejewski}, \citenamefont {Berges},\ and\ \citenamefont
  {Hauke}}]{Zache:2018cqq}%
  \BibitemOpen
  \bibfield  {author} {\bibinfo {author} {\bibfnamefont {T.~V.}\ \bibnamefont
  {Zache}}, \bibinfo {author} {\bibfnamefont {N.}~\bibnamefont {Mueller}},
  \bibinfo {author} {\bibfnamefont {J.~T.}\ \bibnamefont {Schneider}}, \bibinfo
  {author} {\bibfnamefont {F.}~\bibnamefont {Jendrzejewski}}, \bibinfo {author}
  {\bibfnamefont {J.}~\bibnamefont {Berges}}, \ and\ \bibinfo {author}
  {\bibfnamefont {P.}~\bibnamefont {Hauke}},\ }\bibfield  {title} {\enquote
  {\bibinfo {title} {{Dynamical Topological Transitions in the Massive
  Schwinger Model with a $\theta$ Term}},}\ }\href {\doibase
  10.1103/PhysRevLett.122.050403} {\bibfield  {journal} {\bibinfo  {journal}
  {Phys. Rev. Lett.}\ }\textbf {\bibinfo {volume} {122}},\ \bibinfo {pages}
  {050403} (\bibinfo {year} {2019})},\ \Eprint
  {http://arxiv.org/abs/1808.07885} {arXiv:1808.07885 [quant-ph]} \BibitemShut
  {NoStop}%
\bibitem [{\citenamefont {Ikeda}\ \emph {et~al.}(2021)\citenamefont {Ikeda},
  \citenamefont {Kharzeev},\ and\ \citenamefont {Kikuchi}}]{Ikeda:2020agk}%
  \BibitemOpen
  \bibfield  {author} {\bibinfo {author} {\bibfnamefont {Kazuki}\ \bibnamefont
  {Ikeda}}, \bibinfo {author} {\bibfnamefont {Dmitri~E.}\ \bibnamefont
  {Kharzeev}}, \ and\ \bibinfo {author} {\bibfnamefont {Yuta}\ \bibnamefont
  {Kikuchi}},\ }\bibfield  {title} {\enquote {\bibinfo {title} {{Real-time
  dynamics of Chern-Simons fluctuations near a critical point}},}\ }\href
  {\doibase 10.1103/PhysRevD.103.L071502} {\bibfield  {journal} {\bibinfo
  {journal} {Phys. Rev. D}\ }\textbf {\bibinfo {volume} {103}},\ \bibinfo
  {pages} {L071502} (\bibinfo {year} {2021})},\ \Eprint
  {http://arxiv.org/abs/2012.02926} {arXiv:2012.02926 [hep-ph]} \BibitemShut
  {NoStop}%
\bibitem [{\citenamefont {Kharzeev}\ and\ \citenamefont
  {Kikuchi}(2020)}]{Kharzeev:2020kgc}%
  \BibitemOpen
  \bibfield  {author} {\bibinfo {author} {\bibfnamefont {Dmitri~E.}\
  \bibnamefont {Kharzeev}}\ and\ \bibinfo {author} {\bibfnamefont {Yuta}\
  \bibnamefont {Kikuchi}},\ }\bibfield  {title} {\enquote {\bibinfo {title}
  {{Real-time chiral dynamics from a digital quantum simulation}},}\ }\href
  {\doibase 10.1103/PhysRevResearch.2.023342} {\bibfield  {journal} {\bibinfo
  {journal} {Phys. Rev. Res.}\ }\textbf {\bibinfo {volume} {2}},\ \bibinfo
  {pages} {023342} (\bibinfo {year} {2020})},\ \Eprint
  {http://arxiv.org/abs/2001.00698} {arXiv:2001.00698 [hep-ph]} \BibitemShut
  {NoStop}%
\bibitem [{\citenamefont {de~Jong}\ \emph {et~al.}(2022)\citenamefont
  {de~Jong}, \citenamefont {Lee}, \citenamefont {Mulligan}, \citenamefont
  {P\l{}osko\'n}, \citenamefont {Ringer},\ and\ \citenamefont
  {Yao}}]{deJong:2021wsd}%
  \BibitemOpen
  \bibfield  {author} {\bibinfo {author} {\bibfnamefont {Wibe~A.}\ \bibnamefont
  {de~Jong}}, \bibinfo {author} {\bibfnamefont {Kyle}\ \bibnamefont {Lee}},
  \bibinfo {author} {\bibfnamefont {James}\ \bibnamefont {Mulligan}}, \bibinfo
  {author} {\bibfnamefont {Mateusz}\ \bibnamefont {P\l{}osko\'n}}, \bibinfo
  {author} {\bibfnamefont {Felix}\ \bibnamefont {Ringer}}, \ and\ \bibinfo
  {author} {\bibfnamefont {Xiaojun}\ \bibnamefont {Yao}},\ }\bibfield  {title}
  {\enquote {\bibinfo {title} {{Quantum simulation of nonequilibrium dynamics
  and thermalization in the Schwinger model}},}\ }\href {\doibase
  10.1103/PhysRevD.106.054508} {\bibfield  {journal} {\bibinfo  {journal}
  {Phys. Rev. D}\ }\textbf {\bibinfo {volume} {106}},\ \bibinfo {pages}
  {054508} (\bibinfo {year} {2022})},\ \Eprint
  {http://arxiv.org/abs/2106.08394} {arXiv:2106.08394 [quant-ph]} \BibitemShut
  {NoStop}%
\bibitem [{\citenamefont {Xie}\ \emph {et~al.}(2022)\citenamefont {Xie},
  \citenamefont {Guo}, \citenamefont {Xing}, \citenamefont {Xue}, \citenamefont
  {Zhang},\ and\ \citenamefont {Zhu}}]{Xie:2022jgj}%
  \BibitemOpen
  \bibfield  {author} {\bibinfo {author} {\bibfnamefont {Xu-Dan}\ \bibnamefont
  {Xie}}, \bibinfo {author} {\bibfnamefont {Xingyu}\ \bibnamefont {Guo}},
  \bibinfo {author} {\bibfnamefont {Hongxi}\ \bibnamefont {Xing}}, \bibinfo
  {author} {\bibfnamefont {Zheng-Yuan}\ \bibnamefont {Xue}}, \bibinfo {author}
  {\bibfnamefont {Dan-Bo}\ \bibnamefont {Zhang}}, \ and\ \bibinfo {author}
  {\bibfnamefont {Shi-Liang}\ \bibnamefont {Zhu}} (\bibinfo {collaboration}
  {QuNu}),\ }\bibfield  {title} {\enquote {\bibinfo {title} {{Variational
  thermal quantum simulation of the lattice Schwinger model}},}\ }\href
  {\doibase 10.1103/PhysRevD.106.054509} {\bibfield  {journal} {\bibinfo
  {journal} {Phys. Rev. D}\ }\textbf {\bibinfo {volume} {106}},\ \bibinfo
  {pages} {054509} (\bibinfo {year} {2022})},\ \Eprint
  {http://arxiv.org/abs/2205.12767} {arXiv:2205.12767 [quant-ph]} \BibitemShut
  {NoStop}%
\bibitem [{\citenamefont {Belyansky}\ \emph {et~al.}(2024)\citenamefont
  {Belyansky}, \citenamefont {Whitsitt}, \citenamefont {Mueller}, \citenamefont
  {Fahimniya}, \citenamefont {Bennewitz}, \citenamefont {Davoudi},\ and\
  \citenamefont {Gorshkov}}]{Belyansky:2023rgh}%
  \BibitemOpen
  \bibfield  {author} {\bibinfo {author} {\bibfnamefont {Ron}\ \bibnamefont
  {Belyansky}}, \bibinfo {author} {\bibfnamefont {Seth}\ \bibnamefont
  {Whitsitt}}, \bibinfo {author} {\bibfnamefont {Niklas}\ \bibnamefont
  {Mueller}}, \bibinfo {author} {\bibfnamefont {Ali}\ \bibnamefont
  {Fahimniya}}, \bibinfo {author} {\bibfnamefont {Elizabeth~R.}\ \bibnamefont
  {Bennewitz}}, \bibinfo {author} {\bibfnamefont {Zohreh}\ \bibnamefont
  {Davoudi}}, \ and\ \bibinfo {author} {\bibfnamefont {Alexey~V.}\ \bibnamefont
  {Gorshkov}},\ }\bibfield  {title} {\enquote {\bibinfo {title} {{High-Energy
  Collision of Quarks and Mesons in the Schwinger Model: From Tensor Networks
  to Circuit QED}},}\ }\href {\doibase 10.1103/PhysRevLett.132.091903}
  {\bibfield  {journal} {\bibinfo  {journal} {Phys. Rev. Lett.}\ }\textbf
  {\bibinfo {volume} {132}},\ \bibinfo {pages} {091903} (\bibinfo {year}
  {2024})},\ \Eprint {http://arxiv.org/abs/2307.02522} {arXiv:2307.02522
  [quant-ph]} \BibitemShut {NoStop}%
\bibitem [{\citenamefont {Florio}\ \emph {et~al.}(2023)\citenamefont {Florio},
  \citenamefont {Frenklakh}, \citenamefont {Ikeda}, \citenamefont {Kharzeev},
  \citenamefont {Korepin}, \citenamefont {Shi},\ and\ \citenamefont
  {Yu}}]{Florio:2023dke}%
  \BibitemOpen
  \bibfield  {author} {\bibinfo {author} {\bibfnamefont {Adrien}\ \bibnamefont
  {Florio}}, \bibinfo {author} {\bibfnamefont {David}\ \bibnamefont
  {Frenklakh}}, \bibinfo {author} {\bibfnamefont {Kazuki}\ \bibnamefont
  {Ikeda}}, \bibinfo {author} {\bibfnamefont {Dmitri}\ \bibnamefont
  {Kharzeev}}, \bibinfo {author} {\bibfnamefont {Vladimir}\ \bibnamefont
  {Korepin}}, \bibinfo {author} {\bibfnamefont {Shuzhe}\ \bibnamefont {Shi}}, \
  and\ \bibinfo {author} {\bibfnamefont {Kwangmin}\ \bibnamefont {Yu}},\
  }\bibfield  {title} {\enquote {\bibinfo {title} {{Real-Time Nonperturbative
  Dynamics of Jet Production in Schwinger Model: Quantum Entanglement and
  Vacuum Modification}},}\ }\href {\doibase 10.1103/PhysRevLett.131.021902}
  {\bibfield  {journal} {\bibinfo  {journal} {Phys. Rev. Lett.}\ }\textbf
  {\bibinfo {volume} {131}},\ \bibinfo {pages} {021902} (\bibinfo {year}
  {2023})},\ \Eprint {http://arxiv.org/abs/2301.11991} {arXiv:2301.11991
  [hep-ph]} \BibitemShut {NoStop}%
\bibitem [{\citenamefont {Florio}\ \emph {et~al.}(2024)\citenamefont {Florio},
  \citenamefont {Frenklakh}, \citenamefont {Ikeda}, \citenamefont {Kharzeev},
  \citenamefont {Korepin}, \citenamefont {Shi},\ and\ \citenamefont
  {Yu}}]{Florio:2024aix}%
  \BibitemOpen
  \bibfield  {author} {\bibinfo {author} {\bibfnamefont {Adrien}\ \bibnamefont
  {Florio}}, \bibinfo {author} {\bibfnamefont {David}\ \bibnamefont
  {Frenklakh}}, \bibinfo {author} {\bibfnamefont {Kazuki}\ \bibnamefont
  {Ikeda}}, \bibinfo {author} {\bibfnamefont {Dmitri~E.}\ \bibnamefont
  {Kharzeev}}, \bibinfo {author} {\bibfnamefont {Vladimir}\ \bibnamefont
  {Korepin}}, \bibinfo {author} {\bibfnamefont {Shuzhe}\ \bibnamefont {Shi}}, \
  and\ \bibinfo {author} {\bibfnamefont {Kwangmin}\ \bibnamefont {Yu}},\
  }\bibfield  {title} {\enquote {\bibinfo {title} {{Quantum real-time evolution
  of entanglement and hadronization in jet production: Lessons from the massive
  Schwinger model}},}\ }\href {\doibase 10.1103/PhysRevD.110.094029} {\bibfield
   {journal} {\bibinfo  {journal} {Phys. Rev. D}\ }\textbf {\bibinfo {volume}
  {110}},\ \bibinfo {pages} {094029} (\bibinfo {year} {2024})},\ \Eprint
  {http://arxiv.org/abs/2404.00087} {arXiv:2404.00087 [hep-ph]} \BibitemShut
  {NoStop}%
\bibitem [{\citenamefont {Florio}\ \emph {et~al.}(2025)\citenamefont {Florio},
  \citenamefont {Frenklakh}, \citenamefont {Grieninger}, \citenamefont
  {Kharzeev}, \citenamefont {Palermo},\ and\ \citenamefont
  {Shi}}]{Florio:2025hoc}%
  \BibitemOpen
  \bibfield  {author} {\bibinfo {author} {\bibfnamefont {Adrien}\ \bibnamefont
  {Florio}}, \bibinfo {author} {\bibfnamefont {David}\ \bibnamefont
  {Frenklakh}}, \bibinfo {author} {\bibfnamefont {Sebastian}\ \bibnamefont
  {Grieninger}}, \bibinfo {author} {\bibfnamefont {Dmitri~E.}\ \bibnamefont
  {Kharzeev}}, \bibinfo {author} {\bibfnamefont {Andrea}\ \bibnamefont
  {Palermo}}, \ and\ \bibinfo {author} {\bibfnamefont {Shuzhe}\ \bibnamefont
  {Shi}},\ }\bibfield  {title} {\enquote {\bibinfo {title} {{Thermalization
  from quantum entanglement: jet simulations in the massive Schwinger
  model}},}\ }\href@noop {} {\  (\bibinfo {year} {2025})},\ \Eprint
  {http://arxiv.org/abs/2506.14983} {arXiv:2506.14983 [hep-ph]} \BibitemShut
  {NoStop}%
\bibitem [{\citenamefont {Florio}(2024)}]{Florio:2023mzk}%
  \BibitemOpen
  \bibfield  {author} {\bibinfo {author} {\bibfnamefont {Adrien}\ \bibnamefont
  {Florio}},\ }\bibfield  {title} {\enquote {\bibinfo {title} {{Two-fermion
  negativity and confinement in the Schwinger model}},}\ }\href {\doibase
  10.1103/PhysRevD.109.L071501} {\bibfield  {journal} {\bibinfo  {journal}
  {Phys. Rev. D}\ }\textbf {\bibinfo {volume} {109}},\ \bibinfo {pages}
  {L071501} (\bibinfo {year} {2024})},\ \Eprint
  {http://arxiv.org/abs/2312.05298} {arXiv:2312.05298 [hep-th]} \BibitemShut
  {NoStop}%
\bibitem [{\citenamefont {Barata}\ \emph {et~al.}(2024)\citenamefont {Barata},
  \citenamefont {Gong},\ and\ \citenamefont {Venugopalan}}]{Barata:2023jgd}%
  \BibitemOpen
  \bibfield  {author} {\bibinfo {author} {\bibfnamefont {Jo\~ao}\ \bibnamefont
  {Barata}}, \bibinfo {author} {\bibfnamefont {Wenjie}\ \bibnamefont {Gong}}, \
  and\ \bibinfo {author} {\bibfnamefont {Raju}\ \bibnamefont {Venugopalan}},\
  }\bibfield  {title} {\enquote {\bibinfo {title} {{Realtime dynamics of
  hyperon spin correlations from string fragmentation in a deformed four-flavor
  Schwinger model}},}\ }\href {\doibase 10.1103/PhysRevD.109.116003} {\bibfield
   {journal} {\bibinfo  {journal} {Phys. Rev. D}\ }\textbf {\bibinfo {volume}
  {109}},\ \bibinfo {pages} {116003} (\bibinfo {year} {2024})},\ \Eprint
  {http://arxiv.org/abs/2308.13596} {arXiv:2308.13596 [hep-ph]} \BibitemShut
  {NoStop}%
\bibitem [{\citenamefont {Ikeda}\ \emph
  {et~al.}(2023{\natexlab{a}})\citenamefont {Ikeda}, \citenamefont {Kharzeev},
  \citenamefont {Meyer},\ and\ \citenamefont {Shi}}]{Ikeda:2023zil}%
  \BibitemOpen
  \bibfield  {author} {\bibinfo {author} {\bibfnamefont {Kazuki}\ \bibnamefont
  {Ikeda}}, \bibinfo {author} {\bibfnamefont {Dmitri~E.}\ \bibnamefont
  {Kharzeev}}, \bibinfo {author} {\bibfnamefont {Ren\'e}\ \bibnamefont
  {Meyer}}, \ and\ \bibinfo {author} {\bibfnamefont {Shuzhe}\ \bibnamefont
  {Shi}},\ }\bibfield  {title} {\enquote {\bibinfo {title} {{Detecting the
  critical point through entanglement in the Schwinger model}},}\ }\href
  {\doibase 10.1103/PhysRevD.108.L091501} {\bibfield  {journal} {\bibinfo
  {journal} {Phys. Rev. D}\ }\textbf {\bibinfo {volume} {108}},\ \bibinfo
  {pages} {L091501} (\bibinfo {year} {2023}{\natexlab{a}})},\ \Eprint
  {http://arxiv.org/abs/2305.00996} {arXiv:2305.00996 [hep-ph]} \BibitemShut
  {NoStop}%
\bibitem [{\citenamefont {Ikeda}\ \emph
  {et~al.}(2023{\natexlab{b}})\citenamefont {Ikeda}, \citenamefont {Kharzeev},\
  and\ \citenamefont {Shi}}]{Ikeda:2023vfk}%
  \BibitemOpen
  \bibfield  {author} {\bibinfo {author} {\bibfnamefont {Kazuki}\ \bibnamefont
  {Ikeda}}, \bibinfo {author} {\bibfnamefont {Dmitri~E.}\ \bibnamefont
  {Kharzeev}}, \ and\ \bibinfo {author} {\bibfnamefont {Shuzhe}\ \bibnamefont
  {Shi}},\ }\bibfield  {title} {\enquote {\bibinfo {title} {{Nonlinear chiral
  magnetic waves}},}\ }\href {\doibase 10.1103/PhysRevD.108.074001} {\bibfield
  {journal} {\bibinfo  {journal} {Phys. Rev. D}\ }\textbf {\bibinfo {volume}
  {108}},\ \bibinfo {pages} {074001} (\bibinfo {year} {2023}{\natexlab{b}})},\
  \Eprint {http://arxiv.org/abs/2305.05685} {arXiv:2305.05685 [hep-ph]}
  \BibitemShut {NoStop}%
\bibitem [{\citenamefont {Lee}\ \emph {et~al.}(2023)\citenamefont {Lee},
  \citenamefont {Mulligan}, \citenamefont {Ringer},\ and\ \citenamefont
  {Yao}}]{Lee:2023urk}%
  \BibitemOpen
  \bibfield  {author} {\bibinfo {author} {\bibfnamefont {Kyle}\ \bibnamefont
  {Lee}}, \bibinfo {author} {\bibfnamefont {James}\ \bibnamefont {Mulligan}},
  \bibinfo {author} {\bibfnamefont {Felix}\ \bibnamefont {Ringer}}, \ and\
  \bibinfo {author} {\bibfnamefont {Xiaojun}\ \bibnamefont {Yao}},\ }\bibfield
  {title} {\enquote {\bibinfo {title} {{Liouvillian dynamics of the open
  Schwinger model: String breaking and kinetic dissipation in a thermal
  medium}},}\ }\href {\doibase 10.1103/PhysRevD.108.094518} {\bibfield
  {journal} {\bibinfo  {journal} {Phys. Rev. D}\ }\textbf {\bibinfo {volume}
  {108}},\ \bibinfo {pages} {094518} (\bibinfo {year} {2023})},\ \Eprint
  {http://arxiv.org/abs/2308.03878} {arXiv:2308.03878 [quant-ph]} \BibitemShut
  {NoStop}%
\bibitem [{\citenamefont {Ghim}\ and\ \citenamefont
  {Honda}(2024)}]{Ghim:2024pxe}%
  \BibitemOpen
  \bibfield  {author} {\bibinfo {author} {\bibfnamefont {Dongwook}\
  \bibnamefont {Ghim}}\ and\ \bibinfo {author} {\bibfnamefont {Masazumi}\
  \bibnamefont {Honda}},\ }\bibfield  {title} {\enquote {\bibinfo {title}
  {{Digital Quantum Simulation for Spectroscopy of Schwinger Model}},}\ }\href
  {\doibase 10.22323/1.453.0213} {\bibfield  {journal} {\bibinfo  {journal}
  {PoS}\ }\textbf {\bibinfo {volume} {LATTICE2023}},\ \bibinfo {pages} {213}
  (\bibinfo {year} {2024})},\ \Eprint {http://arxiv.org/abs/2404.14788}
  {arXiv:2404.14788 [hep-lat]} \BibitemShut {NoStop}%
\bibitem [{\citenamefont {Li}\ \emph {et~al.}(2022)\citenamefont {Li},
  \citenamefont {Guo}, \citenamefont {Lai}, \citenamefont {Liu}, \citenamefont
  {Wang}, \citenamefont {Xing}, \citenamefont {Zhang},\ and\ \citenamefont
  {Zhu}}]{Li:2021kcs}%
  \BibitemOpen
  \bibfield  {author} {\bibinfo {author} {\bibfnamefont {Tianyin}\ \bibnamefont
  {Li}}, \bibinfo {author} {\bibfnamefont {Xingyu}\ \bibnamefont {Guo}},
  \bibinfo {author} {\bibfnamefont {Wai~Kin}\ \bibnamefont {Lai}}, \bibinfo
  {author} {\bibfnamefont {Xiaohui}\ \bibnamefont {Liu}}, \bibinfo {author}
  {\bibfnamefont {Enke}\ \bibnamefont {Wang}}, \bibinfo {author} {\bibfnamefont
  {Hongxi}\ \bibnamefont {Xing}}, \bibinfo {author} {\bibfnamefont {Dan-Bo}\
  \bibnamefont {Zhang}}, \ and\ \bibinfo {author} {\bibfnamefont {Shi-Liang}\
  \bibnamefont {Zhu}} (\bibinfo {collaboration} {QuNu}),\ }\bibfield  {title}
  {\enquote {\bibinfo {title} {{Partonic collinear structure by quantum
  computing}},}\ }\href {\doibase 10.1103/PhysRevD.105.L111502} {\bibfield
  {journal} {\bibinfo  {journal} {Phys. Rev. D}\ }\textbf {\bibinfo {volume}
  {105}},\ \bibinfo {pages} {L111502} (\bibinfo {year} {2022})},\ \Eprint
  {http://arxiv.org/abs/2106.03865} {arXiv:2106.03865 [hep-ph]} \BibitemShut
  {NoStop}%
\bibitem [{\citenamefont {Li}\ \emph {et~al.}(2023)\citenamefont {Li},
  \citenamefont {Guo}, \citenamefont {Lai}, \citenamefont {Liu}, \citenamefont
  {Wang}, \citenamefont {Xing}, \citenamefont {Zhang},\ and\ \citenamefont
  {Zhu}}]{Li:2022lyt}%
  \BibitemOpen
  \bibfield  {author} {\bibinfo {author} {\bibfnamefont {Tianyin}\ \bibnamefont
  {Li}}, \bibinfo {author} {\bibfnamefont {Xingyu}\ \bibnamefont {Guo}},
  \bibinfo {author} {\bibfnamefont {Wai~Kin}\ \bibnamefont {Lai}}, \bibinfo
  {author} {\bibfnamefont {Xiaohui}\ \bibnamefont {Liu}}, \bibinfo {author}
  {\bibfnamefont {Enke}\ \bibnamefont {Wang}}, \bibinfo {author} {\bibfnamefont
  {Hongxi}\ \bibnamefont {Xing}}, \bibinfo {author} {\bibfnamefont {Dan-Bo}\
  \bibnamefont {Zhang}}, \ and\ \bibinfo {author} {\bibfnamefont {Shi-Liang}\
  \bibnamefont {Zhu}} (\bibinfo {collaboration} {QuNu}),\ }\bibfield  {title}
  {\enquote {\bibinfo {title} {{Exploring light-cone distribution amplitudes
  from quantum computing}},}\ }\href {\doibase 10.1007/s11433-023-2120-1}
  {\bibfield  {journal} {\bibinfo  {journal} {Sci. China Phys. Mech. Astron.}\
  }\textbf {\bibinfo {volume} {66}},\ \bibinfo {pages} {281011} (\bibinfo
  {year} {2023})},\ \Eprint {http://arxiv.org/abs/2207.13258} {arXiv:2207.13258
  [hep-ph]} \BibitemShut {NoStop}%
\bibitem [{\citenamefont {Li}\ \emph {et~al.}(2024)\citenamefont {Li},
  \citenamefont {Lai}, \citenamefont {Wang},\ and\ \citenamefont
  {Xing}}]{Li:2023kex}%
  \BibitemOpen
  \bibfield  {author} {\bibinfo {author} {\bibfnamefont {Tianyin}\ \bibnamefont
  {Li}}, \bibinfo {author} {\bibfnamefont {Wai~Kin}\ \bibnamefont {Lai}},
  \bibinfo {author} {\bibfnamefont {Enke}\ \bibnamefont {Wang}}, \ and\
  \bibinfo {author} {\bibfnamefont {Hongxi}\ \bibnamefont {Xing}} (\bibinfo
  {collaboration} {QuNu}),\ }\bibfield  {title} {\enquote {\bibinfo {title}
  {{Scattering amplitude from quantum computing with reduction formula}},}\
  }\href {\doibase 10.1103/PhysRevD.109.036025} {\bibfield  {journal} {\bibinfo
   {journal} {Phys. Rev. D}\ }\textbf {\bibinfo {volume} {109}},\ \bibinfo
  {pages} {036025} (\bibinfo {year} {2024})},\ \Eprint
  {http://arxiv.org/abs/2301.04179} {arXiv:2301.04179 [hep-ph]} \BibitemShut
  {NoStop}%
\bibitem [{\citenamefont {Czajka}\ \emph {et~al.}(2022)\citenamefont {Czajka},
  \citenamefont {Kang}, \citenamefont {Ma},\ and\ \citenamefont
  {Zhao}}]{Czajka:2021yll}%
  \BibitemOpen
  \bibfield  {author} {\bibinfo {author} {\bibfnamefont {Alexander~M.}\
  \bibnamefont {Czajka}}, \bibinfo {author} {\bibfnamefont {Zhong-Bo}\
  \bibnamefont {Kang}}, \bibinfo {author} {\bibfnamefont {Henry}\ \bibnamefont
  {Ma}}, \ and\ \bibinfo {author} {\bibfnamefont {Fanyi}\ \bibnamefont
  {Zhao}},\ }\bibfield  {title} {\enquote {\bibinfo {title} {{Quantum
  simulation of chiral phase transitions}},}\ }\href {\doibase
  10.1007/JHEP08(2022)209} {\bibfield  {journal} {\bibinfo  {journal} {JHEP}\
  }\textbf {\bibinfo {volume} {08}},\ \bibinfo {pages} {209} (\bibinfo {year}
  {2022})},\ \Eprint {http://arxiv.org/abs/2112.03944} {arXiv:2112.03944
  [hep-ph]} \BibitemShut {NoStop}%
\bibitem [{\citenamefont {Carena}\ \emph {et~al.}(2022)\citenamefont {Carena},
  \citenamefont {Lamm}, \citenamefont {Li},\ and\ \citenamefont
  {Liu}}]{Carena:2022kpg}%
  \BibitemOpen
  \bibfield  {author} {\bibinfo {author} {\bibfnamefont {Marcela}\ \bibnamefont
  {Carena}}, \bibinfo {author} {\bibfnamefont {Henry}\ \bibnamefont {Lamm}},
  \bibinfo {author} {\bibfnamefont {Ying-Ying}\ \bibnamefont {Li}}, \ and\
  \bibinfo {author} {\bibfnamefont {Wanqiang}\ \bibnamefont {Liu}},\ }\bibfield
   {title} {\enquote {\bibinfo {title} {{Improved Hamiltonians for Quantum
  Simulations of Gauge Theories}},}\ }\href {\doibase
  10.1103/PhysRevLett.129.051601} {\bibfield  {journal} {\bibinfo  {journal}
  {Phys. Rev. Lett.}\ }\textbf {\bibinfo {volume} {129}},\ \bibinfo {pages}
  {051601} (\bibinfo {year} {2022})},\ \Eprint
  {http://arxiv.org/abs/2203.02823} {arXiv:2203.02823 [hep-lat]} \BibitemShut
  {NoStop}%
\bibitem [{\citenamefont {Yao}(2023)}]{Yao:2023pht}%
  \BibitemOpen
  \bibfield  {author} {\bibinfo {author} {\bibfnamefont {Xiaojun}\ \bibnamefont
  {Yao}},\ }\bibfield  {title} {\enquote {\bibinfo {title} {{SU(2) gauge theory
  in 2+1 dimensions on a plaquette chain obeys the eigenstate thermalization
  hypothesis}},}\ }\href {\doibase 10.1103/PhysRevD.108.L031504} {\bibfield
  {journal} {\bibinfo  {journal} {Phys. Rev. D}\ }\textbf {\bibinfo {volume}
  {108}},\ \bibinfo {pages} {L031504} (\bibinfo {year} {2023})},\ \Eprint
  {http://arxiv.org/abs/2303.14264} {arXiv:2303.14264 [hep-lat]} \BibitemShut
  {NoStop}%
\bibitem [{\citenamefont {Ebner}\ \emph {et~al.}(2024)\citenamefont {Ebner},
  \citenamefont {M\"uller}, \citenamefont {Sch\"afer}, \citenamefont {Seidl},\
  and\ \citenamefont {Yao}}]{Ebner:2023ixq}%
  \BibitemOpen
  \bibfield  {author} {\bibinfo {author} {\bibfnamefont {Lukas}\ \bibnamefont
  {Ebner}}, \bibinfo {author} {\bibfnamefont {Berndt}\ \bibnamefont
  {M\"uller}}, \bibinfo {author} {\bibfnamefont {Andreas}\ \bibnamefont
  {Sch\"afer}}, \bibinfo {author} {\bibfnamefont {Clemens}\ \bibnamefont
  {Seidl}}, \ and\ \bibinfo {author} {\bibfnamefont {Xiaojun}\ \bibnamefont
  {Yao}},\ }\bibfield  {title} {\enquote {\bibinfo {title} {{Eigenstate
  thermalization in (2+1)-dimensional SU(2) lattice gauge theory}},}\ }\href
  {\doibase 10.1103/PhysRevD.109.014504} {\bibfield  {journal} {\bibinfo
  {journal} {Phys. Rev. D}\ }\textbf {\bibinfo {volume} {109}},\ \bibinfo
  {pages} {014504} (\bibinfo {year} {2024})},\ \Eprint
  {http://arxiv.org/abs/2308.16202} {arXiv:2308.16202 [hep-lat]} \BibitemShut
  {NoStop}%
\bibitem [{\citenamefont {Ikeda}\ \emph {et~al.}(2024)\citenamefont {Ikeda},
  \citenamefont {Kang}, \citenamefont {Kharzeev}, \citenamefont {Qian},\ and\
  \citenamefont {Zhao}}]{Ikeda:2024rzv}%
  \BibitemOpen
  \bibfield  {author} {\bibinfo {author} {\bibfnamefont {Kazuki}\ \bibnamefont
  {Ikeda}}, \bibinfo {author} {\bibfnamefont {Zhong-Bo}\ \bibnamefont {Kang}},
  \bibinfo {author} {\bibfnamefont {Dmitri~E.}\ \bibnamefont {Kharzeev}},
  \bibinfo {author} {\bibfnamefont {Wenyang}\ \bibnamefont {Qian}}, \ and\
  \bibinfo {author} {\bibfnamefont {Fanyi}\ \bibnamefont {Zhao}},\ }\bibfield
  {title} {\enquote {\bibinfo {title} {{Real-time chiral dynamics at finite
  temperature from quantum simulation}},}\ }\href {\doibase
  10.1007/JHEP10(2024)031} {\bibfield  {journal} {\bibinfo  {journal} {JHEP}\
  }\textbf {\bibinfo {volume} {10}},\ \bibinfo {pages} {031} (\bibinfo {year}
  {2024})},\ \Eprint {http://arxiv.org/abs/2407.21496} {arXiv:2407.21496
  [hep-ph]} \BibitemShut {NoStop}%
\bibitem [{\citenamefont {Hayata}\ and\ \citenamefont
  {Hidaka}(2023)}]{Hayata:2023puo}%
  \BibitemOpen
  \bibfield  {author} {\bibinfo {author} {\bibfnamefont {Tomoya}\ \bibnamefont
  {Hayata}}\ and\ \bibinfo {author} {\bibfnamefont {Yoshimasa}\ \bibnamefont
  {Hidaka}},\ }\bibfield  {title} {\enquote {\bibinfo {title} {{String-net
  formulation of Hamiltonian lattice Yang-Mills theories and quantum many-body
  scars in a nonabelian gauge theory}},}\ }\href {\doibase
  10.1007/JHEP09(2023)126} {\bibfield  {journal} {\bibinfo  {journal} {JHEP}\
  }\textbf {\bibinfo {volume} {09}},\ \bibinfo {pages} {126} (\bibinfo {year}
  {2023})},\ \Eprint {http://arxiv.org/abs/2305.05950} {arXiv:2305.05950
  [hep-lat]} \BibitemShut {NoStop}%
\bibitem [{\citenamefont {Hayata}\ \emph {et~al.}(2024)\citenamefont {Hayata},
  \citenamefont {Hidaka},\ and\ \citenamefont {Nishimura}}]{Hayata:2023pkw}%
  \BibitemOpen
  \bibfield  {author} {\bibinfo {author} {\bibfnamefont {Tomoya}\ \bibnamefont
  {Hayata}}, \bibinfo {author} {\bibfnamefont {Yoshimasa}\ \bibnamefont
  {Hidaka}}, \ and\ \bibinfo {author} {\bibfnamefont {Kentaro}\ \bibnamefont
  {Nishimura}},\ }\bibfield  {title} {\enquote {\bibinfo {title} {{Dense
  QCD$_{2}$ with matrix product states}},}\ }\href {\doibase
  10.1007/JHEP07(2024)106} {\bibfield  {journal} {\bibinfo  {journal} {JHEP}\
  }\textbf {\bibinfo {volume} {07}},\ \bibinfo {pages} {106} (\bibinfo {year}
  {2024})},\ \Eprint {http://arxiv.org/abs/2311.11643} {arXiv:2311.11643
  [hep-lat]} \BibitemShut {NoStop}%
\bibitem [{\citenamefont {Hidaka}\ and\ \citenamefont
  {Yamamoto}(2025)}]{Hidaka:2024zkd}%
  \BibitemOpen
  \bibfield  {author} {\bibinfo {author} {\bibfnamefont {Yoshimasa}\
  \bibnamefont {Hidaka}}\ and\ \bibinfo {author} {\bibfnamefont {Arata}\
  \bibnamefont {Yamamoto}},\ }\bibfield  {title} {\enquote {\bibinfo {title}
  {{Quantum circuit for Z3 lattice gauge theory at nonzero baryon density}},}\
  }\href {\doibase 10.1103/PhysRevD.111.014510} {\bibfield  {journal} {\bibinfo
   {journal} {Phys. Rev. D}\ }\textbf {\bibinfo {volume} {111}},\ \bibinfo
  {pages} {014510} (\bibinfo {year} {2025})},\ \Eprint
  {http://arxiv.org/abs/2409.17349} {arXiv:2409.17349 [hep-lat]} \BibitemShut
  {NoStop}%
\bibitem [{\citenamefont {Wu}\ \emph {et~al.}(2024)\citenamefont {Wu},
  \citenamefont {Du}, \citenamefont {Zhao},\ and\ \citenamefont
  {Vary}}]{Wu:2024adk}%
  \BibitemOpen
  \bibfield  {author} {\bibinfo {author} {\bibfnamefont {Sihao}\ \bibnamefont
  {Wu}}, \bibinfo {author} {\bibfnamefont {Weijie}\ \bibnamefont {Du}},
  \bibinfo {author} {\bibfnamefont {Xingbo}\ \bibnamefont {Zhao}}, \ and\
  \bibinfo {author} {\bibfnamefont {James~P.}\ \bibnamefont {Vary}},\
  }\bibfield  {title} {\enquote {\bibinfo {title} {{Efficient and precise
  quantum simulation of ultrarelativistic quark-nucleus scattering}},}\ }\href
  {\doibase 10.1103/PhysRevD.110.056044} {\bibfield  {journal} {\bibinfo
  {journal} {Phys. Rev. D}\ }\textbf {\bibinfo {volume} {110}},\ \bibinfo
  {pages} {056044} (\bibinfo {year} {2024})},\ \Eprint
  {http://arxiv.org/abs/2404.00819} {arXiv:2404.00819 [quant-ph]} \BibitemShut
  {NoStop}%
\bibitem [{\citenamefont {Carena}\ \emph {et~al.}(2024)\citenamefont {Carena},
  \citenamefont {Lamm}, \citenamefont {Li},\ and\ \citenamefont
  {Liu}}]{Carena:2024dzu}%
  \BibitemOpen
  \bibfield  {author} {\bibinfo {author} {\bibfnamefont {Marcela}\ \bibnamefont
  {Carena}}, \bibinfo {author} {\bibfnamefont {Henry}\ \bibnamefont {Lamm}},
  \bibinfo {author} {\bibfnamefont {Ying-Ying}\ \bibnamefont {Li}}, \ and\
  \bibinfo {author} {\bibfnamefont {Wanqiang}\ \bibnamefont {Liu}},\ }\bibfield
   {title} {\enquote {\bibinfo {title} {{Quantum error thresholds for
  gauge-redundant digitizations of lattice field theories}},}\ }\href {\doibase
  10.1103/PhysRevD.110.054516} {\bibfield  {journal} {\bibinfo  {journal}
  {Phys. Rev. D}\ }\textbf {\bibinfo {volume} {110}},\ \bibinfo {pages}
  {054516} (\bibinfo {year} {2024})},\ \Eprint
  {http://arxiv.org/abs/2402.16780} {arXiv:2402.16780 [hep-lat]} \BibitemShut
  {NoStop}%
\bibitem [{\citenamefont {Qian}\ \emph {et~al.}(2025)\citenamefont {Qian},
  \citenamefont {Li}, \citenamefont {Salgado},\ and\ \citenamefont
  {Kreshchuk}}]{Qian:2024gph}%
  \BibitemOpen
  \bibfield  {author} {\bibinfo {author} {\bibfnamefont {Wenyang}\ \bibnamefont
  {Qian}}, \bibinfo {author} {\bibfnamefont {Meijian}\ \bibnamefont {Li}},
  \bibinfo {author} {\bibfnamefont {Carlos~A.}\ \bibnamefont {Salgado}}, \ and\
  \bibinfo {author} {\bibfnamefont {Michael}\ \bibnamefont {Kreshchuk}},\
  }\bibfield  {title} {\enquote {\bibinfo {title} {{Efficient quantum
  simulation of QCD jets on the light front}},}\ }\href {\doibase
  10.1103/PhysRevD.111.096001} {\bibfield  {journal} {\bibinfo  {journal}
  {Phys. Rev. D}\ }\textbf {\bibinfo {volume} {111}},\ \bibinfo {pages}
  {096001} (\bibinfo {year} {2025})},\ \Eprint
  {http://arxiv.org/abs/2411.09762} {arXiv:2411.09762 [hep-ph]} \BibitemShut
  {NoStop}%
\bibitem [{\citenamefont {Janik}\ \emph {et~al.}(2025)\citenamefont {Janik},
  \citenamefont {Nowak}, \citenamefont {Rams},\ and\ \citenamefont
  {Zahed}}]{Janik:2025bbz}%
  \BibitemOpen
  \bibfield  {author} {\bibinfo {author} {\bibfnamefont {Romuald~A.}\
  \bibnamefont {Janik}}, \bibinfo {author} {\bibfnamefont {Maciej~A.}\
  \bibnamefont {Nowak}}, \bibinfo {author} {\bibfnamefont {Marek~M.}\
  \bibnamefont {Rams}}, \ and\ \bibinfo {author} {\bibfnamefont {Ismail}\
  \bibnamefont {Zahed}},\ }\bibfield  {title} {\enquote {\bibinfo {title}
  {{Universality and emergent effective fluid from jets and string breaking in
  the massive Schwinger model using tensor networks}},}\ }\href@noop {} {\
  (\bibinfo {year} {2025})},\ \Eprint {http://arxiv.org/abs/2502.12901}
  {arXiv:2502.12901 [hep-ph]} \BibitemShut {NoStop}%
\bibitem [{\citenamefont {Chen}\ \emph {et~al.}(2024)\citenamefont {Chen},
  \citenamefont {Yan},\ and\ \citenamefont {Shi}}]{Chen:2024pee}%
  \BibitemOpen
  \bibfield  {author} {\bibinfo {author} {\bibfnamefont {Shile}\ \bibnamefont
  {Chen}}, \bibinfo {author} {\bibfnamefont {Li}~\bibnamefont {Yan}}, \ and\
  \bibinfo {author} {\bibfnamefont {Shuzhe}\ \bibnamefont {Shi}},\ }\bibfield
  {title} {\enquote {\bibinfo {title} {{Quantum thermalization of Quark-Gluon
  Plasma}},}\ }\href@noop {} {\  (\bibinfo {year} {2024})},\ \Eprint
  {http://arxiv.org/abs/2412.00662} {arXiv:2412.00662 [hep-ph]} \BibitemShut
  {NoStop}%
\bibitem [{\citenamefont {Turro}\ and\ \citenamefont
  {Yao}(2025)}]{Turro:2025sec}%
  \BibitemOpen
  \bibfield  {author} {\bibinfo {author} {\bibfnamefont {Francesco}\
  \bibnamefont {Turro}}\ and\ \bibinfo {author} {\bibfnamefont {Xiaojun}\
  \bibnamefont {Yao}},\ }\bibfield  {title} {\enquote {\bibinfo {title}
  {{Emergent hydrodynamic mode on SU(2) plaquette chains and quantum
  simulation}},}\ }\href {\doibase 10.1103/PhysRevD.111.094502} {\bibfield
  {journal} {\bibinfo  {journal} {Phys. Rev. D}\ }\textbf {\bibinfo {volume}
  {111}},\ \bibinfo {pages} {094502} (\bibinfo {year} {2025})},\ \Eprint
  {http://arxiv.org/abs/2502.17551} {arXiv:2502.17551 [hep-ph]} \BibitemShut
  {NoStop}%
\bibitem [{\citenamefont {Bauer}\ \emph
  {et~al.}(2023{\natexlab{a}})\citenamefont {Bauer} \emph
  {et~al.}}]{Bauer:2022hpo}%
  \BibitemOpen
  \bibfield  {author} {\bibinfo {author} {\bibfnamefont {Christian~W.}\
  \bibnamefont {Bauer}} \emph {et~al.},\ }\bibfield  {title} {\enquote
  {\bibinfo {title} {{Quantum Simulation for High-Energy Physics}},}\ }\href
  {\doibase 10.1103/PRXQuantum.4.027001} {\bibfield  {journal} {\bibinfo
  {journal} {PRX Quantum}\ }\textbf {\bibinfo {volume} {4}},\ \bibinfo {pages}
  {027001} (\bibinfo {year} {2023}{\natexlab{a}})},\ \Eprint
  {http://arxiv.org/abs/2204.03381} {arXiv:2204.03381 [quant-ph]} \BibitemShut
  {NoStop}%
\bibitem [{\citenamefont {Bauer}\ \emph
  {et~al.}(2023{\natexlab{b}})\citenamefont {Bauer}, \citenamefont {Davoudi},
  \citenamefont {Klco},\ and\ \citenamefont {Savage}}]{Bauer:2023qgm}%
  \BibitemOpen
  \bibfield  {author} {\bibinfo {author} {\bibfnamefont {Christian~W.}\
  \bibnamefont {Bauer}}, \bibinfo {author} {\bibfnamefont {Zohreh}\
  \bibnamefont {Davoudi}}, \bibinfo {author} {\bibfnamefont {Natalie}\
  \bibnamefont {Klco}}, \ and\ \bibinfo {author} {\bibfnamefont {Martin~J.}\
  \bibnamefont {Savage}},\ }\bibfield  {title} {\enquote {\bibinfo {title}
  {{Quantum simulation of fundamental particles and forces}},}\ }\href
  {\doibase 10.1038/s42254-023-00599-8} {\bibfield  {journal} {\bibinfo
  {journal} {Nature Rev. Phys.}\ }\textbf {\bibinfo {volume} {5}},\ \bibinfo
  {pages} {420--432} (\bibinfo {year} {2023}{\natexlab{b}})},\ \Eprint
  {http://arxiv.org/abs/2404.06298} {arXiv:2404.06298 [hep-ph]} \BibitemShut
  {NoStop}%
\bibitem [{\citenamefont {Rommer}\ and\ \citenamefont
  {Ostlund}(1997)}]{Rommer:1997zz}%
  \BibitemOpen
  \bibfield  {author} {\bibinfo {author} {\bibfnamefont {Stefan}\ \bibnamefont
  {Rommer}}\ and\ \bibinfo {author} {\bibfnamefont {Stellan}\ \bibnamefont
  {Ostlund}},\ }\bibfield  {title} {\enquote {\bibinfo {title} {{Class of
  ansatz wave functions for one-dimensional spin systems and their relation to
  the density matrix renormalization group}},}\ }\href {\doibase
  10.1103/PhysRevB.55.2164} {\bibfield  {journal} {\bibinfo  {journal} {Phys.
  Rev. B}\ }\textbf {\bibinfo {volume} {55}},\ \bibinfo {pages} {2164--2181}
  (\bibinfo {year} {1997})},\ \Eprint {http://arxiv.org/abs/cond-mat/9606213}
  {arXiv:cond-mat/9606213} \BibitemShut {NoStop}%
\bibitem [{\citenamefont {Buyens}\ \emph {et~al.}(2014)\citenamefont {Buyens},
  \citenamefont {Haegeman}, \citenamefont {Van~Acoleyen}, \citenamefont
  {Verschelde},\ and\ \citenamefont {Verstraete}}]{Buyens:2013yza}%
  \BibitemOpen
  \bibfield  {author} {\bibinfo {author} {\bibfnamefont {Boye}\ \bibnamefont
  {Buyens}}, \bibinfo {author} {\bibfnamefont {Jutho}\ \bibnamefont
  {Haegeman}}, \bibinfo {author} {\bibfnamefont {Karel}\ \bibnamefont
  {Van~Acoleyen}}, \bibinfo {author} {\bibfnamefont {Henri}\ \bibnamefont
  {Verschelde}}, \ and\ \bibinfo {author} {\bibfnamefont {Frank}\ \bibnamefont
  {Verstraete}},\ }\bibfield  {title} {\enquote {\bibinfo {title} {{Matrix
  product states for gauge field theories}},}\ }\href {\doibase
  10.1103/PhysRevLett.113.091601} {\bibfield  {journal} {\bibinfo  {journal}
  {Phys. Rev. Lett.}\ }\textbf {\bibinfo {volume} {113}},\ \bibinfo {pages}
  {091601} (\bibinfo {year} {2014})},\ \Eprint {http://arxiv.org/abs/1312.6654}
  {arXiv:1312.6654 [hep-lat]} \BibitemShut {NoStop}%
\bibitem [{\citenamefont {Buyens}\ \emph {et~al.}(2016)\citenamefont {Buyens},
  \citenamefont {Verstraete},\ and\ \citenamefont
  {Van~Acoleyen}}]{Buyens:2016ecr}%
  \BibitemOpen
  \bibfield  {author} {\bibinfo {author} {\bibfnamefont {Boye}\ \bibnamefont
  {Buyens}}, \bibinfo {author} {\bibfnamefont {Frank}\ \bibnamefont
  {Verstraete}}, \ and\ \bibinfo {author} {\bibfnamefont {Karel}\ \bibnamefont
  {Van~Acoleyen}},\ }\bibfield  {title} {\enquote {\bibinfo {title}
  {{Hamiltonian simulation of the Schwinger model at finite temperature}},}\
  }\href {\doibase 10.1103/PhysRevD.94.085018} {\bibfield  {journal} {\bibinfo
  {journal} {Phys. Rev. D}\ }\textbf {\bibinfo {volume} {94}},\ \bibinfo
  {pages} {085018} (\bibinfo {year} {2016})},\ \Eprint
  {http://arxiv.org/abs/1606.03385} {arXiv:1606.03385 [hep-lat]} \BibitemShut
  {NoStop}%
\bibitem [{\citenamefont {Kogut}\ and\ \citenamefont
  {Susskind}(1975)}]{Kogut:1974ag}%
  \BibitemOpen
  \bibfield  {author} {\bibinfo {author} {\bibfnamefont {John~B.}\ \bibnamefont
  {Kogut}}\ and\ \bibinfo {author} {\bibfnamefont {Leonard}\ \bibnamefont
  {Susskind}},\ }\bibfield  {title} {\enquote {\bibinfo {title} {{Hamiltonian
  Formulation of Wilson's Lattice Gauge Theories}},}\ }\href {\doibase
  10.1103/PhysRevD.11.395} {\bibfield  {journal} {\bibinfo  {journal} {Phys.
  Rev. D}\ }\textbf {\bibinfo {volume} {11}},\ \bibinfo {pages} {395--408}
  (\bibinfo {year} {1975})}\BibitemShut {NoStop}%
\bibitem [{\citenamefont {Susskind}(1977)}]{Susskind:1976jm}%
  \BibitemOpen
  \bibfield  {author} {\bibinfo {author} {\bibfnamefont {Leonard}\ \bibnamefont
  {Susskind}},\ }\bibfield  {title} {\enquote {\bibinfo {title} {{Lattice
  Fermions}},}\ }\href {\doibase 10.1103/PhysRevD.16.3031} {\bibfield
  {journal} {\bibinfo  {journal} {Phys. Rev. D}\ }\textbf {\bibinfo {volume}
  {16}},\ \bibinfo {pages} {3031--3039} (\bibinfo {year} {1977})}\BibitemShut
  {NoStop}%
\bibitem [{\citenamefont {Brenes}\ \emph {et~al.}(2018)\citenamefont {Brenes},
  \citenamefont {Dalmonte}, \citenamefont {Heyl},\ and\ \citenamefont
  {Scardicchio}}]{Brenes:2017wzd}%
  \BibitemOpen
  \bibfield  {author} {\bibinfo {author} {\bibfnamefont {Marlon}\ \bibnamefont
  {Brenes}}, \bibinfo {author} {\bibfnamefont {Marcello}\ \bibnamefont
  {Dalmonte}}, \bibinfo {author} {\bibfnamefont {Markus}\ \bibnamefont {Heyl}},
  \ and\ \bibinfo {author} {\bibfnamefont {Antonello}\ \bibnamefont
  {Scardicchio}},\ }\bibfield  {title} {\enquote {\bibinfo {title} {{Many-body
  localization dynamics from gauge invariance}},}\ }\href {\doibase
  10.1103/PhysRevLett.120.030601} {\bibfield  {journal} {\bibinfo  {journal}
  {Phys. Rev. Lett.}\ }\textbf {\bibinfo {volume} {120}},\ \bibinfo {pages}
  {030601} (\bibinfo {year} {2018})},\ \Eprint
  {http://arxiv.org/abs/1706.05878} {arXiv:1706.05878 [cond-mat.str-el]}
  \BibitemShut {NoStop}%
\bibitem [{\citenamefont {Vidal}(2004)}]{Vidal:2003lvx}%
  \BibitemOpen
  \bibfield  {author} {\bibinfo {author} {\bibfnamefont {G.}~\bibnamefont
  {Vidal}},\ }\bibfield  {title} {\enquote {\bibinfo {title} {{Efficient
  simulation of one-dimensional quantum many-body systems}},}\ }\href {\doibase
  10.1103/PhysRevLett.93.040502} {\bibfield  {journal} {\bibinfo  {journal}
  {Phys. Rev. Lett.}\ }\textbf {\bibinfo {volume} {93}},\ \bibinfo {pages}
  {040502} (\bibinfo {year} {2004})},\ \Eprint
  {http://arxiv.org/abs/quant-ph/0310089} {arXiv:quant-ph/0310089} \BibitemShut
  {NoStop}%
\bibitem [{\citenamefont {Shao}\ \emph {et~al.}(2025)\citenamefont {Shao},
  \citenamefont {Chen},\ and\ \citenamefont {Shi}}]{Shao_2025_long}%
  \BibitemOpen
  \bibfield  {author} {\bibinfo {author} {\bibfnamefont {Haiyang}\ \bibnamefont
  {Shao}}, \bibinfo {author} {\bibfnamefont {Shile}\ \bibnamefont {Chen}}, \
  and\ \bibinfo {author} {\bibfnamefont {Shuzhe}\ \bibnamefont {Shi}},\
  }\bibfield  {title} {\enquote {\bibinfo {title} {{Collective motion in the
  massive Schwinger model via Tensor Network}},}\ }\href@noop {} {\  (\bibinfo
  {year} {2025})},\ \Eprint {http://arxiv.org/abs/2509.xxxxx} {arXiv:2509.xxxxx
  [hep-ph]} \BibitemShut {NoStop}%
\bibitem [{\citenamefont {Hwa}(1974)}]{Hwa:1974gn}%
  \BibitemOpen
  \bibfield  {author} {\bibinfo {author} {\bibfnamefont {Rudolph~C.}\
  \bibnamefont {Hwa}},\ }\bibfield  {title} {\enquote {\bibinfo {title}
  {{Statistical Description of Hadron Constituents as a Basis for the Fluid
  Model of High-Energy Collisions}},}\ }\href {\doibase
  10.1103/PhysRevD.10.2260} {\bibfield  {journal} {\bibinfo  {journal} {Phys.
  Rev. D}\ }\textbf {\bibinfo {volume} {10}},\ \bibinfo {pages} {2260}
  (\bibinfo {year} {1974})}\BibitemShut {NoStop}%
\bibitem [{\citenamefont {Bjorken}(1983)}]{Bjorken:1982qr}%
  \BibitemOpen
  \bibfield  {author} {\bibinfo {author} {\bibfnamefont {J.~D.}\ \bibnamefont
  {Bjorken}},\ }\bibfield  {title} {\enquote {\bibinfo {title} {{Highly
  Relativistic Nucleus-Nucleus Collisions: The Central Rapidity Region}},}\
  }\href {\doibase 10.1103/PhysRevD.27.140} {\bibfield  {journal} {\bibinfo
  {journal} {Phys. Rev. D}\ }\textbf {\bibinfo {volume} {27}},\ \bibinfo
  {pages} {140--151} (\bibinfo {year} {1983})}\BibitemShut {NoStop}%
\bibitem [{\citenamefont {Shi}\ \emph {et~al.}(2022)\citenamefont {Shi},
  \citenamefont {Jeon},\ and\ \citenamefont {Gale}}]{Shi:2022iyb}%
  \BibitemOpen
  \bibfield  {author} {\bibinfo {author} {\bibfnamefont {Shuzhe}\ \bibnamefont
  {Shi}}, \bibinfo {author} {\bibfnamefont {Sangyong}\ \bibnamefont {Jeon}}, \
  and\ \bibinfo {author} {\bibfnamefont {Charles}\ \bibnamefont {Gale}},\
  }\bibfield  {title} {\enquote {\bibinfo {title} {{Family of new exact
  solutions for longitudinally expanding ideal fluids}},}\ }\href {\doibase
  10.1103/PhysRevC.105.L021902} {\bibfield  {journal} {\bibinfo  {journal}
  {Phys. Rev. C}\ }\textbf {\bibinfo {volume} {105}},\ \bibinfo {pages}
  {L021902} (\bibinfo {year} {2022})},\ \Eprint
  {http://arxiv.org/abs/2201.06670} {arXiv:2201.06670 [hep-ph]} \BibitemShut
  {NoStop}%
\end{thebibliography}%
\end{document}